\documentclass[conference]{IEEEtran}
\IEEEoverridecommandlockouts
\usepackage{cite}
\usepackage{amsmath,amssymb,amsfonts}
\usepackage{graphicx}
\usepackage{textcomp}
\usepackage{xcolor}
\def\BibTeX{{\rm B\kern-.05em{\sc i\kern-.025em b}\kern-.08em
    T\kern-.1667em\lower.7ex\hbox{E}\kern-.125emX}}  
\usepackage{array}
\usepackage{textcomp}
\usepackage{stfloats}
\usepackage{url}
\usepackage{verbatim}
\usepackage{enumitem}
\usepackage{algpseudocode}
\usepackage[bottom]{footmisc}
\usepackage{mathtools}
\usepackage{subcaption}
\def\BibTeX{{\rm B\kern-.05em{\sc i\kern-.025em b}\kern-.08em
    T\kern-.1667em\lower.7ex\hbox{E}\kern-.125emX}}
\DeclarePairedDelimiter\floor{\lfloor}{\rfloor}
\usepackage[linesnumbered,ruled,vlined]{algorithm2e}
\algrenewcommand{\algorithmiccomment}[1]{\hfill// #1}
\algnewcommand{\algorithmicor}{\textbf{ or }}
\algnewcommand{\OR}{\algorithmicor}
\usepackage{balance}
\begin{document}
\title{GT-TSCH: Game-Theoretic Distributed TSCH Scheduler for Low-Power IoT Networks}

\author{

\IEEEauthorblockN{ Omid Tavallaie\textsuperscript{1}, Seid Miad Zandavi\textsuperscript{2}, Hamed Haddadi\textsuperscript{3}, and Albert Y. Zomaya\textsuperscript{1}}
\IEEEauthorblockA{\textsuperscript{1}School of Computer Science, The University of Sydney, Australia\\
\textsuperscript{2} The Broad Institute of MIT and Harvard, USA\\
\textsuperscript{3} Imperial College London, UK\\
\{omid.tavallaie, albert.zomaya\}@sydney.edu.au, szandavi@broadinstitute.org, h.haddadi@imperial.ac.uk}}

\maketitle

\begin{abstract}
Time-Slotted Channel Hopping (TSCH) is a synchronous medium access mode of the IEEE 802.15.4e standard designed for providing low-latency and highly-reliable end-to-end communication. TSCH constructs a communication schedule by combining frequency channel hopping with Time Division Multiple Access (TDMA). In recent years, IETF designed several standards to define general mechanisms for the implementation of TSCH. However, the problem of updating the TSCH schedule according to the changes of the wireless link quality and node's traffic load left unresolved. In this paper, we use non-cooperative game theory to propose GT-TSCH, a distributed TSCH scheduler designed for low-power IoT applications. By considering selfish behavior of nodes in packet forwarding, GT-TSCH updates the TSCH schedule in a distributed approach with low control overhead by monitoring the queue length, the place of the node in the Directed Acyclic Graph (DAG) topology, the quality of the wireless link, and the data packet generation rate. We prove the existence and uniqueness of Nash equilibrium in our game model and we find the optimal number of TSCH Tx timeslots to update the TSCH slotframe. To examine the performance of our contribution, we implement GT-TSCH on Zolertia Firefly IoT motes and the Contiki-NG Operating System (OS). The evaluation results reveal that GT-TSCH improves performance in terms of throughput and end-to-end delay compared to the state-of-the-art method.
\end{abstract}

\section{Introduction}
Flourishing versatile applications of low-power Internet of Things (IoT) has surged the demand for exploring high-throughput and energy-efficient communication protocols. From environment monitoring to smart cities and mission critical applications, low-power resource-constrained devices use multi-hop communication over short-range wireless links to create low-power IoT networks which are main building blocks of Cyber-Physical Systems (CPS). Due to resource scarcity of embedded devices, lightweight asynchronous link layer protocols were employed at the early stage of low-power IoT networks, which are entirely different from conventional protocols used in traditional networks such as TCP/IP. This trend has been changed in recent years by receiving  enormous attention from industry \cite{rfc5867, rfc5826, rfc5673}, increasing traffic demand, and developing hardware and network technologies.
\begin{figure}[b]\vspace{-4 mm}
 \centering
 \includegraphics[width= 84 mm, height=40 mm]{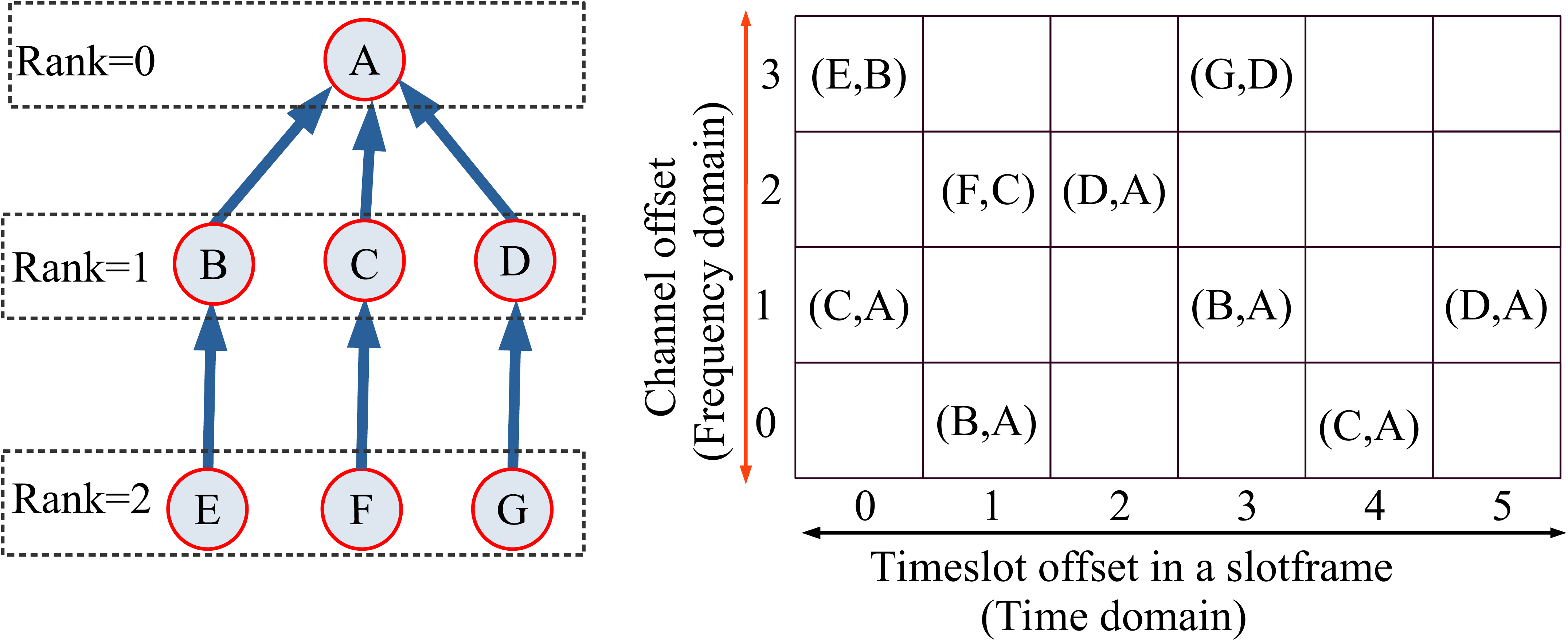} 
  \caption{A CDU matrix with 4 available frequency channels when the size of the slotframe is 6 timeslots. Filling a cell with the node pair (B,A) indicates sending a packet from node B to node A.}
  \label{fig:1}
\end{figure}
Since 2012, IETF and IEEE have been designing various standardized protocols \cite{mac, rfc6550, rfc6551, rfc6552, rfc6554, rfc8480, rfc8180} to connect resource-constrained devices to the Internet. Recently, IEEE standardized Time-Slotted Channel Hopping (TSCH) as a synchronous Medium Access Control (MAC) mode of IEEE 802.15.4e \cite{mac}. By combining multi frequency channels, Time Division Multiple Access (TDMA), and frequency channel hopping, TSCH provides low-latency and reliable communication over lossy wireless links in mesh networks. It cuts time into a fixed number of \textit{timeslots} called \textit{slotframe}. The duration of each timeslot is long enough (around 10 to 15 milliseconds) for sending a packet to an one-hop neighbor placed in the wireless range of an IoT node and receiving an acknowledgement (ACK). 

To create a schedule for distributing all communications into two-dimensional space (time and frequency), TSCH constructs a Channel Distribution Usage (CDU) matrix. Each cell of the CDU matrix is addressed by a pair of (timeslot offset, frequency offset) and it is used for transmission of one packet between adjacent IoT nodes. Fig. \ref{fig:1} shows an example of a CDU matrix for a Directed Acyclic Graph (DAG) topology constructed by using the hop-count metric as the objective function of the RPL routing protocol \cite{rfc6550}. 

To bind TSCH with protocols designed for Low-power Lossy Networks (LLNs), the IETF 6TiSCH Working Group \cite{6tisch} has been designing standards since 2015. The 6TiSCH  protocol \cite{rfc8180} defines a sublayer named 6top (6P) \cite{rfc8480} for integrating TSCH into the protocol stack of LLNs. The 6top sublayer enables neighbor nodes to update their TSCH schedules based on a Scheduling Function (SF). As stated in RFC 8180 \cite{rfc8180} and RFC 8480 \cite{rfc8480}, dynamic SFs that could be self-adjusted according to time-varying traffic are left as open research problems. At each TSCH timeslot, an IoT node can transmit/receive a packet or stay in the sleep mode to save energy. As the TSCH SF determines the node's action (transmit/receive/sleep), it heavily impacts the node's traffic load and its radio duty cycle. 

In this paper, we introduce GT-TSCH, a distributed adaptive TSCH scheduling function designed based on the non-cooperative game theory which is widely used in communication systems for modeling and analyzing interactions between network nodes \cite{game, mara, jacob, ipsn}. GT-TSCH tightly interacts with the RPL routing protocol to dynamically adjust the TSCH slotframe of each IoT node in a distributed fashion with low control overhead. To avoid congestion and balance the traffic load in the network, GT-TSCH adjusts the TSCH schedule by monitoring 1) the node's queue length, 2) the quality of the wireless link, 3) the place of the node in the DAG topology, 4) and the traffic generation rate. Main contributions of this paper are as follows:

\begin{itemize}[topsep=0 pt, partopsep=0 pt, wide=0 pt]
\item GT-TSCH is designed based on the non-cooperative game theory optimization. To find an optimal solution for the problem of creating and updating the TSCH schedule, we model the process of TSCH timeslot allocation as a non-cooperative game where each IoT node selfishly tries to allocate more TSCH Tx timeslots for maximizing its data generation rate. First, we prove the existence and uniqueness of Nash equilibrium based on the Rosen's theorem \cite{rosen} for concave N-person games. Then, we model finding the optimal solution as a nonlinear programming problem.

\item Most of the well-known TSCH schedulers (e.g. \cite{orchestra, alice, a3}) do not have any mechanism to consider wireless interference. They assign frequency channels randomly to different nodes by using a hash function. In our proposed method, we can limit wireless interference and collisions significantly for nodes placed on the same Destination Oriented Directed Acyclic Graph (DODAG) topology. 

\item GT-TSCH avoids congestion and node failure by considering link quality, the node’s Rank, and the queue length in the game model. We use Expected Transmission Count (ETX) metric to take wireless link quality into account in the cost function of our game model. By degrading the wireless link quality, we increase the cost in the payoff function to reduce the number of TSCH Tx timeslots for updating the slotframe.

\item We implement GT-TSCH on Contiki-NG \cite{contiki} Operating System (OS) and Zolertia Firefly IoT motes \cite{Zolertia} that have limited 32 KB of RAM. GT-TSCH is fully compatible with other protocols designed for LLNs. 
\end{itemize}

The remainder of this paper is organized as follows: Section \ref{sectionrelated} discusses related work. Section \ref{channel} explains the channel allocation process of GT-TSCH by using an example. Section \ref{slotframe} details the GT-TSCH's slotframe creation process. Section \ref{allocation} explains how GT-TSCH allocates unicast data timeslots in a distributed fashion. Section \ref{load} introduces the GT-TSCH's load balancing algorithm. Section \ref{game} models the problem of updating the TSCH schedule as a non-cooperative game and finds the optimal solution. Section \ref{evaluation} is about implementing GT-TSCH on Contiki-NG OS, evaluating its performance, and analyzing the results. Finally, section \ref{conclusion} concludes this paper. Table {\ref{table:1}} shows the main symbols used in this paper. 
\begin{table}[b]\vspace{-2 mm}
\renewcommand{\arraystretch}{1.2}
\caption{Main symbols and definitions}
\label{table:1}
\centering
\begin{tabular}{|*{55}{p{10 mm}|p{68 mm}|}}
\hline
\textbf{\hspace{-1 mm}Symbol}  &\textbf{Definition}\\ \hline \hline
$N$  &Set of all IoT nodes\\ \hline
$p_{i}$  &Parent of node $i$ in the DAG topology\\ \hline
$F$  &Set of frequency channels\\ \hline
$f_{i,j}$  &Channel used for sending packets from node $i$ to node $j$\\  \hline
$f_{bcast}$  &Channel used for broadcasting control packets\\  \hline
$cs_{i}$   &Children set of node $i$ in the DAG topology\\  \hline
$l_{i}^{rx}$ &Number of unicast reception cells of node $i$\\  \hline
$l_{i}^{tx}$ &Number of unicast transmission cells of node $i$\\  \hline
$l_{i}^{g}$ &Number of cells used for packet generation at node $i$\\  \hline
$Rank_{i}$ &Rank of node $i$ in the DAG topology\\  \hline
$ETX_{i,p_{i}}$ & Expected transmission count metric for the link $(i,p_{i})$ \\  \hline
$m$   &Size of the TSCH slotframe\\  \hline
$q_{i}$  &Queue length of node $i$\\  \hline
$Q_{Max}$  &Maximum queue length\\  \hline
$Q_{i}$  &Weighted average queue metric of node $i$\\  \hline 
\end{tabular} 
\end{table}
\section{Related Work} \label{sectionrelated}
\subsection{Distributed TSCH Schedulers}
Jung \textit{et al}. proposed SSAP in \cite{tsch:ssap} for providing Quality of Service (QoS) guarantee in TSCH networks. SSAP is designed in a distributed fashion to maximize the network lifetime and satisfy reliability and latency requirements. Regarding energy maximization, SSAP employs the alpha-fairness theory to instill the fairness of lifetime among nodes. \cite{oa-tsch} and \cite{dtsf} propose solutions for optimizing the slotframe creation process based on the convex optimization theory. DT-SF \cite{dtsf} is a distributed TSCH scheduler designed for mobile applications of low-power IoT networks. By using a lightweight approach, DT-SF estimates the node mobility based on the number of parent changes and the duration of a connection with the parent node. DT-SF models timeslot allocation as a mixed-integer convex optimization problem. In \cite{fly}, Wang \textit{et al}. proposed HF-OTF, a hysteresis-free on-the-fly scheduling function. HF-OTF avoids congestion and determines the required bandwidth without considering the configuration of a hysteresis quantum which is the application-specific threshold for network resources in the cell allocation policies.
 
\cite{ldsf} and \cite{ysf} propose distributed SFs for reducing delay in real-time applications. \cite{ldsf} divides the slotframe into small blocks. On each node, a block that minimizes the delay to the border router is selected. \cite{ysf} is designed for multipoint-to-point traffic. It considers network dynamic metrics including the network formation phase and packet switching in the scheduling function to minimize the latency and maximize reliability. REA-6TiSCH \cite{rea} is a reliable communication scheme designed for supporting emergency alarms in 6TiSCH networks. REA-6TiSCH hijacks transmission cells preassigned to regular traffic for the emergency traffic. In addition, REA-6TiSCH employs a distributed optimization scheme for enhancing the probability of delivering traffic before a deadline.

\subsection{Centralized TSCH Schedulers}
\cite{tsch:max} addresses the problems of fairness and throughput maximization in centralized TSCH schedules. The problem of throughput maximization was formulated as an integer programming problem solved by a proposed algorithm with the polynomial time. Ojo \textit{et al}. proposed an energy-efficient scheduler and a heuristic scheduling algorithm in \cite{ees} based on the Vogel’s approximation method. They formulated the energy efficiency maximization of TSCH scheduling as a nonlinear integer programming problem. To decrease the computational complexity of the solution, a greedy-based energy-efficient scheduler was proposed that assigns a frequency channel to a node to maximize its energy efficiency. \cite{white} proposes a solution for creating a whitelist of frequency channels for centralized TSCH schedulers to improve reliability. To avoid collisions, all the communications scheduled for the same timeslot are forced to use the same whitelist. Besides, \cite{white} proposes an algorithm that reorders the whitelists to forbid any possible collisions. 
\subsection{Autonomous TSCH Schedulers}
In \cite{tsch:tmc19}, Vallati \textit{et al}. assessed the network formation dynamic of 6TiSCH networks and showed that the resource allocation of 6TiSCH minimal configuration could cause a significant delay in the network formation, which may lead to a disconnected network topology. In \cite{orchestra}, Duquennoy \textit{et al}. proposed Orchestra, an autonomous scheduler for creating a robust mesh network. For each traffic plane, Orchestra maintains a schedule and updates it automatically when the topology changes. Without increasing control packet overhead, Orchestra employs network stack information to build local schedules. It enables each node to compute its own schedule by using information of one-hop neighbors. Alice, a link-based autonomous scheduling method that allocates a unique cell for each directional link in the DAG topology, was introduced in \cite{alice}. Alice allocates more TSCH cells to nodes with a higher number of one-hop neighbors. It assigns different cells to a node for upstream and downstream traffic. TESLA \cite{tsch:tesla} is a traffic-aware elastic slotframe adjustment scheme that enables each node in the TSCH network to adjust the slotframe size dynamically at run time to minimize energy consumption. TESLA adjusts the schedule for receiving packets by considering traffic load. When the contention is high, TESLA decreases the size of the slotframe to improve the throughput. When the contention is low, the slotframe size is increased to save energy resources.

\section{Channel Allocation Process}\label{channel}
To create the CDU matrix, channel and timeslot offsets must be determined by the TSCH scheduler for communication of each node with its parent and its children. Channel/timeslot allocation policies considerably impact the packet delivery ratio of wireless links. To design an effective TSCH scheduling function for low-power IoT networks, the impact of these policies on root causes of wireless interference and collision should be investigated.   

Most of the well-known TSCH schedulers (e.g., \cite{orchestra,a3,ost,alice}) select channel and timeslot offsets by using hash functions. Through extensive experiments, we found four cases where these methods cause severe wireless interference problems. In this section, we use an example in Fig. {\ref{fig:problem}} to explain these problems. \textbf{1) Using the same timeslot offset for communication of a node with both of its parent and its children}: The duration of each timeslot is around 15 milliseconds which is enough for sending a packet to an one-hop neighbor and receiving the ACK. Hence, each node can have only one packet transmission/reception in a timeslot. In Fig. {\ref{fig:problem}}, the communication of node B with both of nodes A and E (red edges) are scheduled for timeslot 1. This issue incurs collisions on concurrent packet transmissions (B,A) and (E,B). \textbf{2) Using the same frequency channel for communication of sibling nodes with their children}: In the DAG topology, the radio coverage of sibling nodes usually overlap. This problem is shown in Fig. {\ref{fig:problem}} for nodes D and C. In the CDU matrix, communication of these nodes with their children (purple edges) are scheduled for the TSCH cell (4,1). These concurrent transmissions are collided as CSMA/CA is not used for the  transmission of data packets. Note that using TSCH cells with different timeslots cannot alleviate this problem in dense networks. As an example, 25 nodes can be placed in the DAG topology with the maximum distance of two hops from the root node (border router) when the number of children is 5. Thus, for any slotframe with the size of less than 25 timeslots, at least two unicast packet transmissions are scheduled for the same TSCH cell. {\textbf{3) Assigning the same frequency channel to two nodes when their distances to the root node differ by one hop:} Similar to the problem 2, nodes are usually placed inside the wireless ranges of their uncles in the DAG topology. This problem is shown in Fig. {\ref{fig:problem}} for edges (D,A) and (G,C) (colored by orange). This problem can be solved by avoid allocating the same channel to two nodes where the difference between their distances from the root node is one hop. {\textbf{4) Assigning the same channel to two nodes with two-hop distance from each other:} As shown for edges (L,G) and (C,A) (green color) in Fig. {\ref{fig:problem}}, using the same channel for nodes with two hop distance may cause the hidden terminal problem \cite{hiddenTerminal}. Node G is placed inside the communication ranges of both nodes C and L. Hence, collisions occurs on concurrent unicast transmissions of (L,G) and (C,A). This problem cannot be  alleviated by scheduling these transmissions on different timeslots (similar to problem 2). To address these 4 major problems, GT-TSCH allocates frequency channels based on the following strategies:
\begin{itemize}[topsep=0 pt, partopsep=0 pt, wide=0 pt]
\item GT-TSCH schedules the communication of a parent node with its children by using only one channel. As the parent node can receive just one packet per timeslot, each timeslot is assigned to only one child node. Fig. {\ref{fig:freq2}} shows the result of implementing this strategy on the topology of Fig. {\ref{fig:problem}}. As shown in Fig. {\ref{fig:freq2}}, all children nodes use the same frequency channel for forwarding data packets to the parent node. For instance, node A receives traffic from all of its children (nodes B, C, and D) on channel $f1$.
\item GT-TSCH assigns two different channels to each node in the network for communication with its parent and children (as shown in Fig. {\ref{fig:freq2}}). The channel that node $i$ can use for forwarding data to its parent $p_{i}$, is piggybacked on TSCH Enhanced Beacon (EB) messages which are broadcast periodically by $p_{i}$. To inform a node of the channel for communication with its children, we create a new command code \textit{ASK-CHANNEL} for 6P advertisement messages (Fig. {\ref{fig:6p}}). By receiving an EB message from $p_{i}$, node $i$ sends an \textit{ASK-CHANNEL-REQUEST} message to $p_{i}$ to know the channel $f_{i,cs_{i}}$ that connects node $i$ to its children ($cs_{i}$). As an example, in Fig. {\ref{fig:freq2}}, node A informs its children of channel $f_{1}$ by broadcasting an EB message. Then, node B (or nodes C and D) sends an \textit{ASK-CHANNEL-REQUEST} message to node A to know the channel it can use for receiving data from its children (nodes E and F). When node A replies this request, it informs node B of channel $f_{2}$ stored in the \textit{Channel Offset} field of the response message.  
\begin{figure}[t]
	\centering 
    \includegraphics[width=88 mm, height= 40 mm]{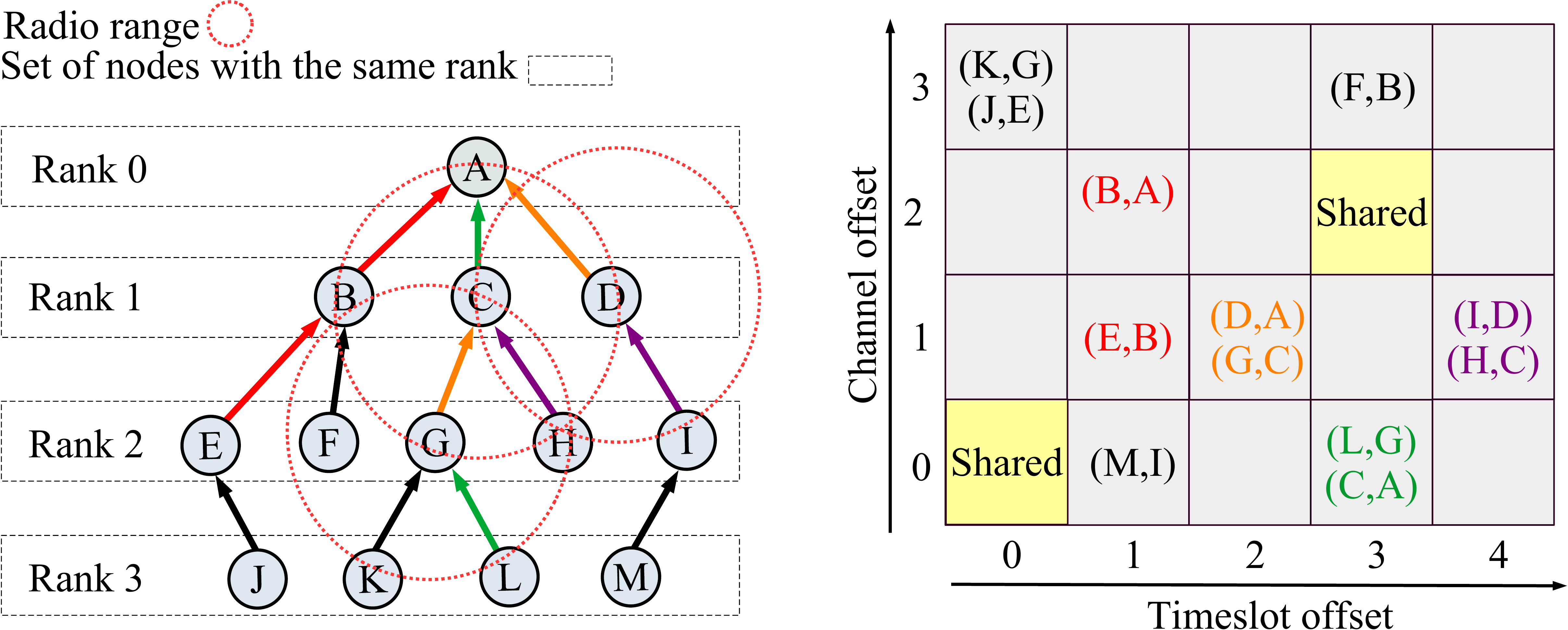}\vspace{-1 mm}
    \caption{The impact of TSCH timeslot allocation policies on collisions and wireless interference.}
\label{fig:problem}
\end{figure}

\item To cope with problem 4, GT-TSCH keeps each allocated channel unique on three-hop routing paths. To implement this strategy, we limit the number of children in the DAG topology. For $n$ available channels, GT-TSCH uses one channel $f_{bcast}$ for broadcasting control packets and limit the number of children to $n-2-1$ to make each channel different with those used at previous and next hops of the routing path. For example, as shown in Fig. {\ref{fig:freq2}}, the channel used for the communication (G,C) is different with those ones assigned to (C,A) and (K,G). To implement this strategy, GT-TSCH responds the \textit{ASK-CHANNEL-REQUEST} message by selecting channels that have not been used in the next two hops (towards the root node). For instance, node $i$ responds the \textit{ASK-CHANNEL-REQUEST} message by selecting a channel $f\in F$ where $f \notin  \{f_{i,cs_{i}},f_{i,p_{i}},f_{bcast}\}$. Algorithm \ref{alg:1} shows the GT-TSCH's channel allocation process running at node $i$.
\end{itemize} 
\begin{figure}[t]
	\centering 
    \includegraphics[width=60 mm, height= 40 mm]{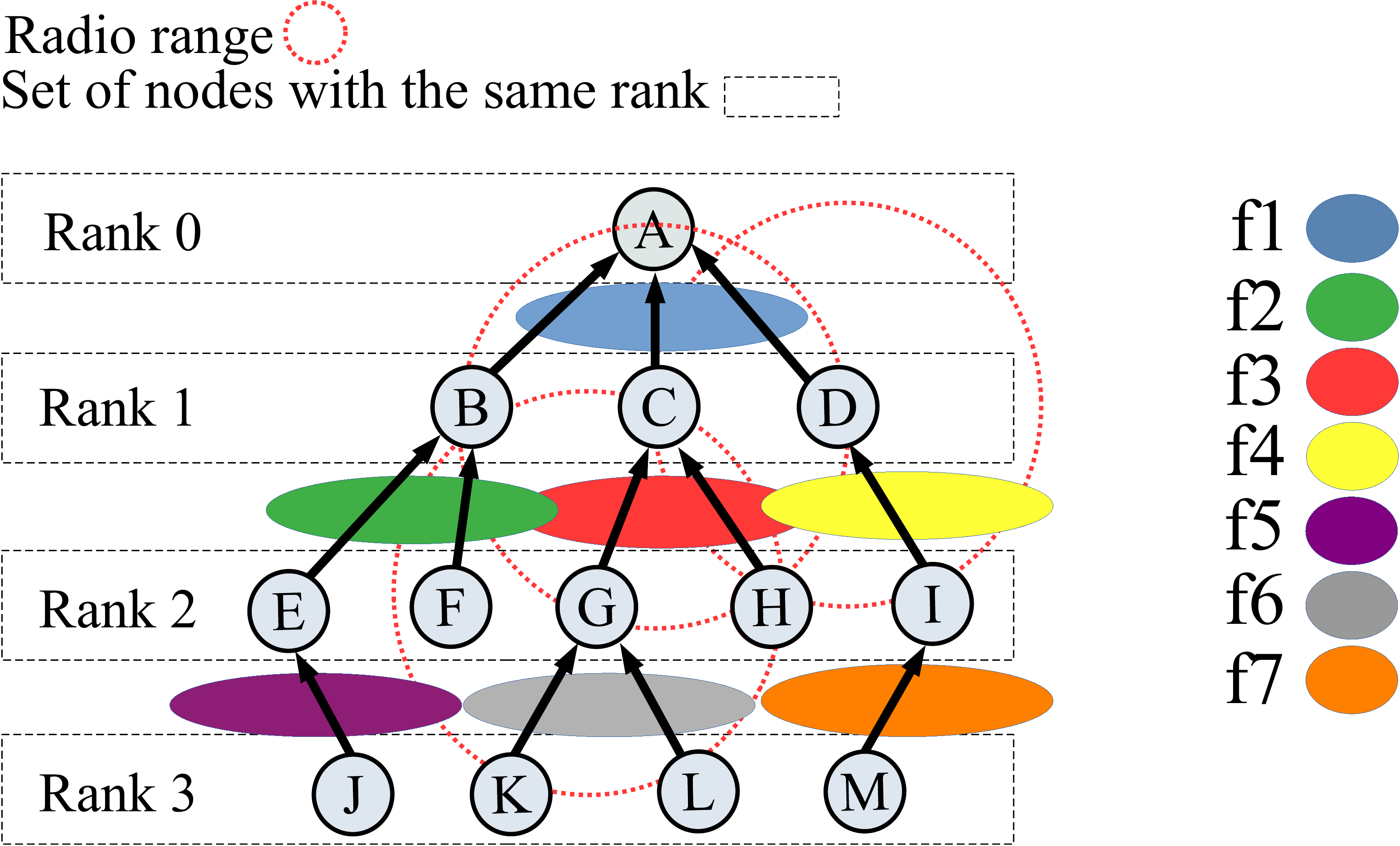}
    \caption{The result of running the GT-TSCH's channel allocation algorithm with 7 available channels.}
    \label{fig:freq2}
\end{figure}  

\begin{figure}[b] 
  \centering
  \begin{subfigure}{1\linewidth}
    \centering
    \includegraphics[width=70 mm , height= 15 mm]{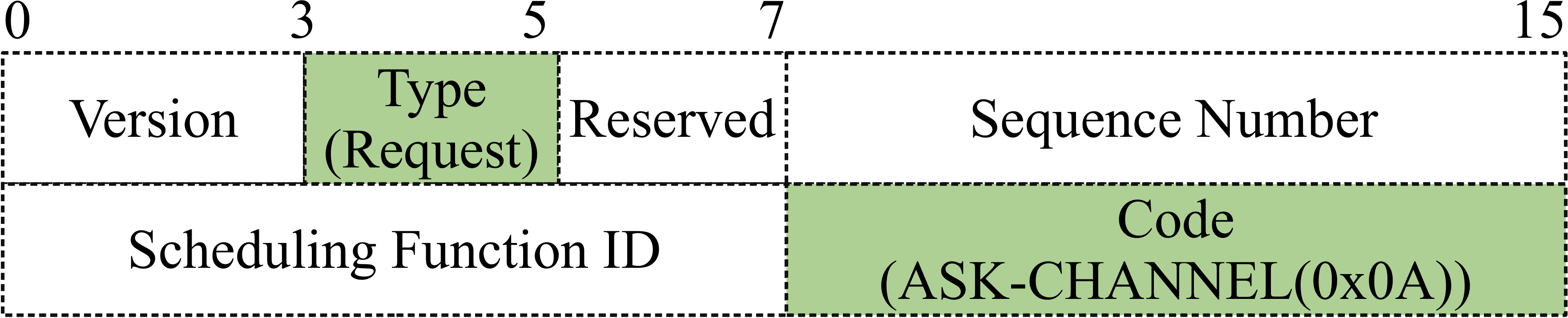}
    \label{fig:6p1} 
    \caption{6P \textit{ASK-CHANNEL-REQUEST} message format.} \vspace{2 mm}
  \end{subfigure}
  \begin{subfigure}{1\linewidth}
     \centering
     \includegraphics[width=70mm , height=19 mm]{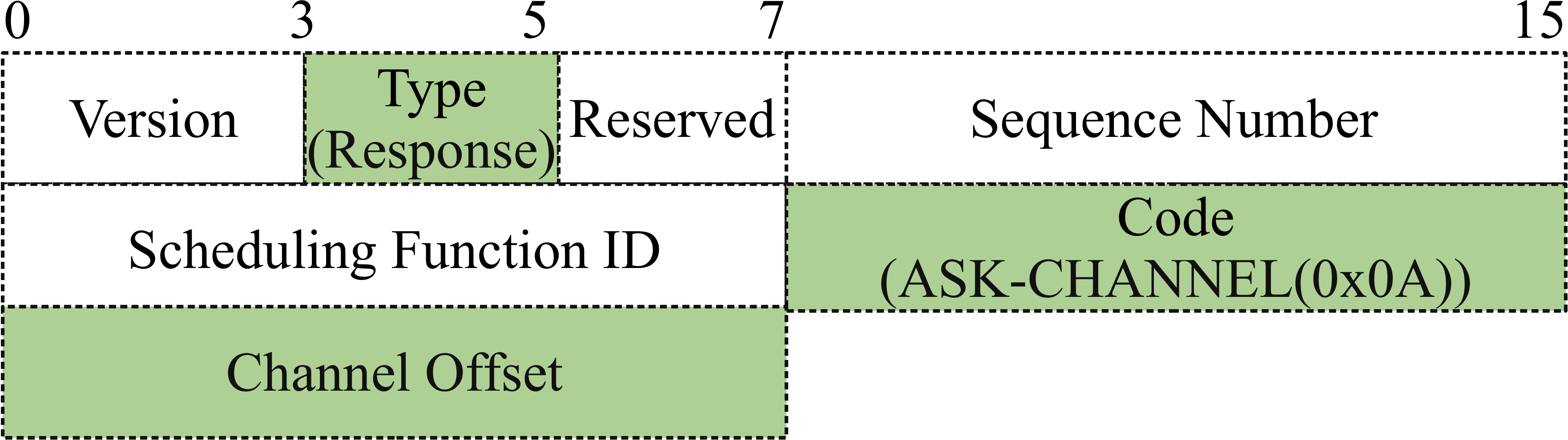}
     \label{fig:6p2}
     \caption{6P \textit{ASK-CHANNEL-RESPONSE} message format.} 
  \end{subfigure}
  \caption{Format of 6P \textit{ASK-CHANNEL} messages used to inform a node of the channel allocated for the communication with the children.}
  \label{fig:6p}
\end{figure}
\begin{algorithm}[t]
$•$
\SetAlgoLined
\DontPrintSemicolon
$f_{i,p_{i}} \gets NULL \And f_{i,cs_{i}} \gets NULL$\newline
\uIf{$i==root$}
{
\text{// Select a random channel from the set $F-\{f_{bcast}\}$}
$f_{i,cs_{i}} \gets Rand(F-\{f_{bcast}\})$ 
\text{// Root nodes do not have parents}\newline
}
\Else
{
  \While{$f_{i,p_{i}}==NULL  \OR f_{i,cs_{i}}==NULL$}{
\text{Receive values of $f_{i,p_{i}}$ and $f_{i,cs_{i}}$ from } 
\text{TSCH EB and 6P \textit{ASK-CHANNEL} messages}  
  }

}

\For{$ \forall k \in cs_{i}$}  
 {
 	$f_{k,cs_{k}} \gets NULL$
 }
 \For{$\forall j \in cs_{i}$}{
 \For{$\forall z \in F-\{f_{bcast},f_{i,p_{i}},f_{i,cs_{i}}\}$}  
 {             
  	  {\vspace{1 mm}
  	  	  $check \gets 0$ \newline
          \For{$ \forall l \in cs_{i},\; l \neq j$}  
 		  {\vspace{1 mm}
 		  	\If{$z == f_{l,cs_{l}}$}
  	  		{\vspace{1 mm}
  	  		 $check \gets 1$ \newline
  	  		 \textbf{Break}
  	  		}
 		  	
 		  }
  	  	 
 		  \If{$check == 0$}
  	  	  {\vspace{1 mm}
  	  		 $f_{j,cs_{j}} \gets z$ \vspace{1 mm}\newline
  	  		 \textbf{Break}
  	  	  }
 		
  	  }
  	  }
 }
\caption{The GT-TSCH's channel allocation process running at node $i$ for finding $f_{i,p_{i}}, f_{i,cs_{i}}, \text{and } f_{j,cs_{j}(\forall j \in cs_{i})}$. } \label{alg:1}
\end{algorithm}\setlength{\textfloatsep}{3 mm}
\section{The Slotframe Creation Process}\label{slotframe}
GT-TSCH defines five types of timeslots which are sorted by the descending order of priority as follows:
\begin{itemize}[topsep=0 pt, partopsep=0 pt, wide=0 pt]
\item \textbf{Broadcast} timeslots are employed for broadcasting control packets (RPL/TSCH). According to the size of the slotframe, GT-TSCH allocates a fixed number of broadcast timeslots. 
 \item \textbf{Unicast-6P} timeslots are employed for updating the TSCH schedule through unicast transmissions of 6P \textit{ADD/DELETE} messages exchanged between two adjacent neighbors. 
 \item \textbf{Unicast-Data} timeslots are allocated for forwarding data packets from a child node to its parent in the DAG topology.
\item \textbf{Shared} timeslots are assigned to a node and its children for unicast transmission of data packets. GT-TSCH uses shared timeslots to eliminate the impact of sudden changes of data traffic on the node's load. Back-off algorithms are employed in these timeslots for resolving contention.
\item \textbf{Sleep} timeslots have the least priority in the slotframe. In these timeslots, nodes turn off their radio transmitters to save energy resources. Sleep mode is the default type of all timeslots when the slotframe is initialized.
\end{itemize}
By taking these five types of timeslots into account, GT-TSCH creates the slotframe on each node based on following rules: \textbf{1)} The highest priority is considered for broadcast timeslots as they are used for building the DAG topology and creating the TSCH schedule. In the initialization process of the TSCH operation, GT-TSCH creates a slotframe with the size of $m \in \mathbb{N}^+$ with $k \in \mathbb{N}^+(k<m)$ broadcast timeslots for sending/receiving control packets. Values of $m$ and $k$ are set based on the numbers of roots and IoT nodes in the network. GT-TSCH uniformly distributes broadcast timeslots in the slotframe to shorten the time interval between each two consecutive timeslots. This is done by allocating broadcast timeslots with offsets $j=\{x|x \in \mathbb{N}^{0}, x<m,x\% \floor{m/k}=0\}$ where $\%$ is the modulo operation. As an example when $m=20, k=5$, $j=\{0,4,8,12,16\}$. \textbf{2)} After creating the DAG topology, each node sends a 6P \textit{ADD-REQUEST} message to its parent for allocating  Unicast-6P timeslots. These timeslots are used for updating the TSCH schedule by exchanging 6P \textit{ADD/DELETE} messages. To provide a highly reliable framework for updating the TSCH schedule, GT-TSCH allocates a fixed number of Unicast-6P timeslots based on the size of the slotframe. For instance, in our experiments, we allocate two Unicast-6P timeslots for the communication between two nodes when the size of the slotframe is 32. Hence, for a node with five children and one parent, the total number of allocated Unicast-6P timeslots per slotframe is computed as $(5+1)\times 2=12$. \textbf{3)} For adding/removing Unicast-Data timeslots, nodes send 6P \textit{ADD/DELETE} messages at Unicast-6P timeslots. GT-TSCH monitors the node's load periodically to update the number of allocated Unicast-Data timeslots. When the traffic load is light, GT-TSCH decreases the number of Unicast-Data timeslots to save energy resources by turning off the radio transmitter. Under heavy traffic, GT-TSCH allocates more Unicast-Data timeslots to reduce the node's load. \textbf{4)} To eliminate the impact of sudden changes of data traffic on the node's load, a fix number of timeslots are shared between children and the parent node for unicast transmission of data/6P packets. We use CSMA/CA in these timeslots to resolve contention. In our experiments, we set the number of shared timeslots equal to the half of the maximum number of children in the DAG topology. Hence, each shared timeslot is used by two children nodes for sending/receiving a packet to/from the parent node.
\begin{figure}[b]
      \begin{subfigure}{1\linewidth}
      	\centering
		\includegraphics[width=70 mm, height= 20 mm]{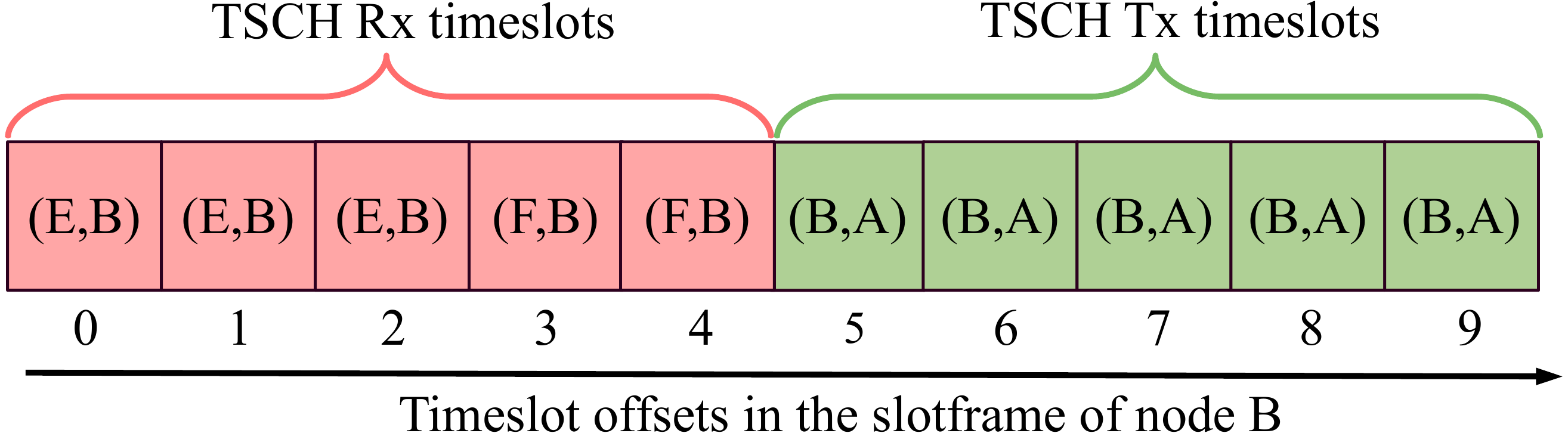} 		  		\label{fig:congestion-a}
		\caption{}
		 \label{fig:congestion-a}
		\vspace{2 mm}
  	  \end{subfigure}
  	  \begin{subfigure}[b]{1\linewidth}
  	  	 \centering
 		 \includegraphics[width=70 mm, height=14 mm]{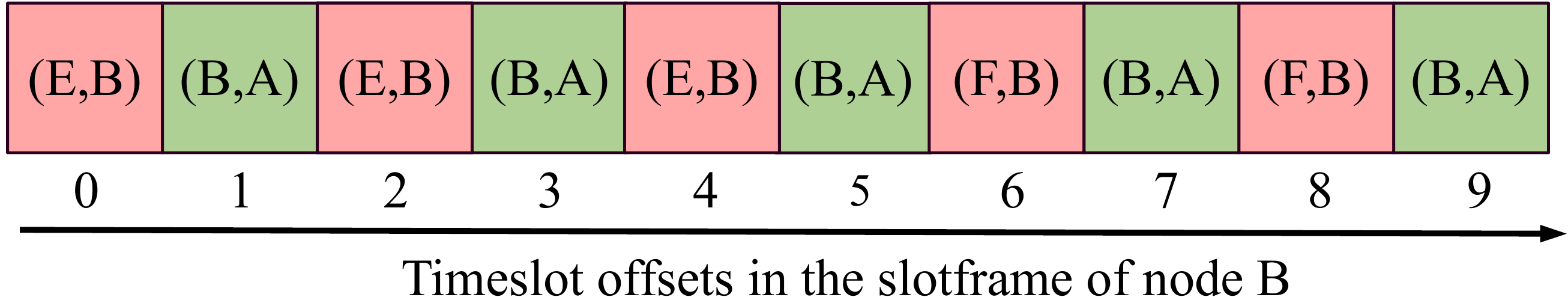}	
 		 \caption{}
 		 \label{fig:congestion-b}
 		 \vspace{2 mm}
  	  \end{subfigure}
      \begin{subfigure}[c]{1\linewidth}
         \centering
 		 \includegraphics[width=70 mm, height=14 mm]{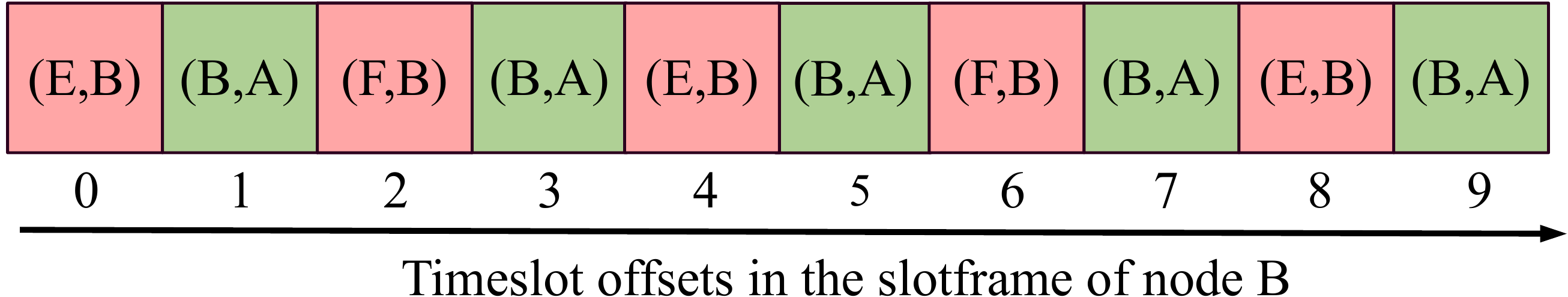}
 		 \caption{}\vspace{-1 mm}
 		 \label{fig:congestion-c}
  	  \end{subfigure}
  \caption{The impact of allocating consecutive Rx timeslots on the node's load.}
  \label{fig:congestion}
\end{figure}
\section{The Allocation Process for Unicast-Data Timeslots}\label{allocation}
To balance the node's load and distributes the traffic load in the DAG topology, GT-TSCH applies the following rules for assigning Unicast-Data timeslots to children nodes:
\begin{itemize}[wide=0pt]
\item GT-TSCH keeps the number of TSCH Tx timeslots always higher than the number of TSCH Rx timeslots per slotframe. This policy ensures that the total incoming traffic rate of each node during a slotframe is less than its total outgoing rate.

\item To decrease the number of queued packets, GT-TSCH allocates at least one TSCH Tx timeslot between two consecutive TSCH Rx timeslots. Without considering this strategy, allocation of TSCH Rx timeslots may lead to a significant increase in the queue length. Fig. {\ref{fig:congestion-a}} shows a case when this problem causes full congestion at node B in Fig. {\ref{fig:problem}} where the slotframe size is 10. Before timeslots $t5$, node B has to keep all the received packets in the queue as there is no TSCH Tx timeslot for forwarding them. By considering the maximum queue length of 4, the node B's queue is getting full at the end of timeslot 3 and it has to drop a packet at timeslot 4. This problem can be easily solved by allocating at least one TSCH Tx timeslot in the slotframe between each two consecutive TSCH Rx timeslots, as it is shown in Fig. {\ref{fig:congestion-b}}.

\item To reduce the end-to-end delay, GT-TSCH minimizes the time that a packet spends at each node in the routing path by implementing a fair strategy for distribution of TSCH Rx timeslots between children nodes. In Fig.  {\ref{fig:congestion-b}}, out of five TSCH Rx timeslots allocated in the slotframe, the last two timeslots are assigned to node F. Thus, when node F is ready to forward a data packet at the beginning of the slotframe, it needs to wait for at least 6$\times$15 milliseconds (in just one hop of the routing path). To solve this problem, GT-TSCH avoids assigning two consecutive TSCH Rx timeslots to a child node as long as there is more than one node with data packets ready for transmission. Fig. \ref{fig:7} shows the result of running the GT-TSCH's timeslot allocation process in a network of 7 nodes.\vspace{-1 mm}
\end{itemize}

\begin{figure*}[t]
	\centering 
    \includegraphics[width=165 mm, height= 60 mm]{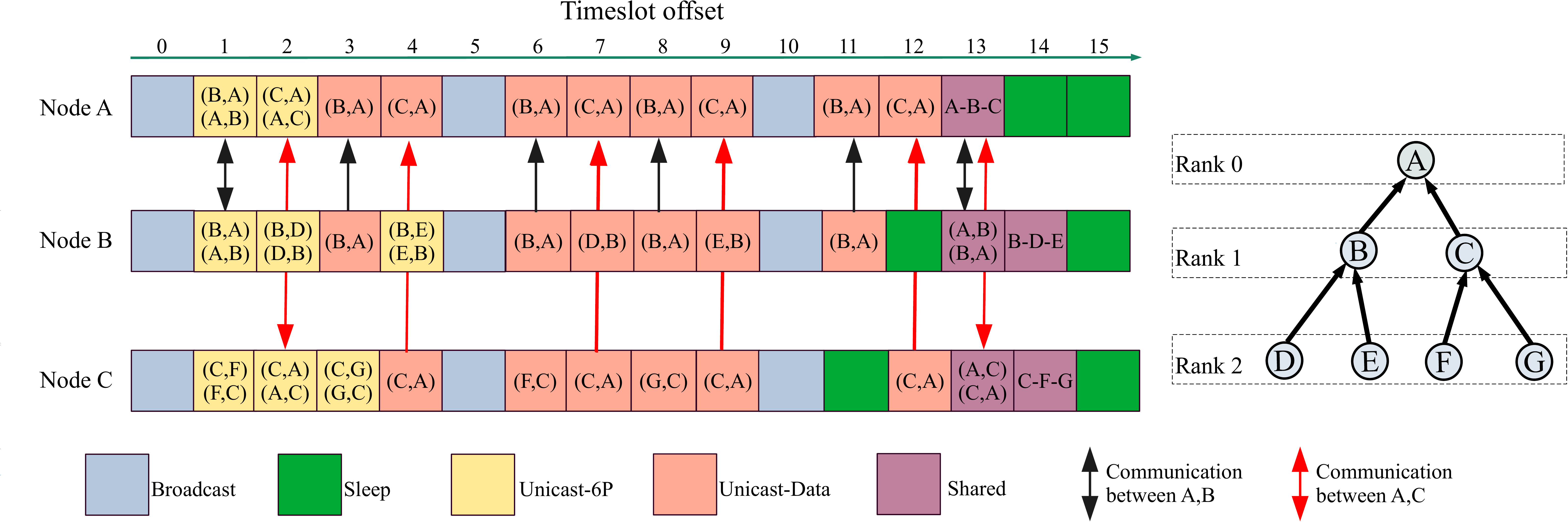}
    \caption{The result of running the GT-TSCH's timeslot allocation process in a DAG network topology of 7 nodes.}
    \label{fig:7}\vspace{-2 mm}
\end{figure*}

\section{Load Balancing}\label{load}
Changing the numbers of TSCH Tx and TSCH Rx timeslots directly impact the packet forwarding and receiving rates. To balance the node's load and avoid congestion, the TSCH scheduler should be updated when the receiving or forwarding traffic rates are changed. More TSCH Tx timeslots should be allocated when the packet generation rate of hosted applications or packet receiving rates are increased. In this case, a node sends a 6P \textit{ADD-REQUEST} message to its parent node for allocating new TSCH Tx timeslots when the current number of free TSCH Tx timeslots is less than the sum of 1) the number TSCH Tx timeslots required for the packet generation and 2) the total number of TSCH Rx timeslots requested from children nodes. For updating the TSCH schedule, GT-TSCH monitors the rates of packet generation of hosted applications periodically and computes the minimum number of required Tx cells at node $i$ by \vspace{-1 mm}
\begin{equation}\label{eq:3.1}
l^{tx-min}_{i}= l^{g}_{i} + l^{tx}_{cs_{i}}-l^{tx-free}_{i},
\end{equation}
where $l^{g}_{i}$ is the number of TSCH Tx timeslots required for the packet generation, $r^{tx}_{cs_{i}}$ is the total number of requested TSCH Tx timeslots that node $i$ receives in 6P \textit{ADD-REQUEST} messages from its children, and $l^{tx-free}_{i}$ is the current number of free TSCH Tx timeslots allocated in node i's slotframe. By computing $l^{tx-min}_{i}$ periodically, node $i$ updates its TSCH schedule to balance its traffic load. While $l_{i}^{tx-min}$ shows the minimum number of required TSCH Tx timeslots, a node can request allocation a higher number of TSCH Tx timeslots to 1) tolerate the sudden increase in the packet receiving rate, and 2) increase the packet generation rate of hosted applications. By considering nodes' selfish behavior in maximizing throughput, a node may request allocation of the maximum number TSCH Tx cells. in the next subsection, we model finding the value of $l^{tx}_{i}$ as a non-cooperative game.
\section{Game Model and Problem Formulation}\label{game}\vspace{1 mm}
In a network with a set of $n$ IoT nodes ${N=\{1,2,3,...,n\}}$, we model the problem of allocating TSCH cells as a non-cooperative game $G= (N,(S_{i})i \in N, (v_{i}) i \in N )$ with $n$ players (IoT nodes). ${v_{i}}$ and ${S_{i}}$ are the payoff function and the strategy set of player $i$, respectively. In this game, players compete for receiving more TSCH Tx cells to forward data packets to root nodes of the DAG topology. By using a timer, each node $i \in N$ runs the load balancing algorithm periodically and sends 6P \textit{ADD-REQUEST} messages to its parent node $p_{i}$ when $l_{p_{i}}^{rx}>0$ and $l^{tx-min}_{i}>0$. The number of requested TSCH Tx cells ($l_{i}^{tx}$) is set equal to $l^{rx}_{p_{i}}$ when $l^{rx}_{p_{i}} \le l^{tx-min}_{i}$. Else, node $i$ selects a strategy $s_{i} \in S_{i}$ where $S_{i}=[l^{tx-min}_{i},l^{rx}_{p_{i}}]$ is the strategy set for selecting the value of $l_{i}^{tx}$. Hence, ${S=\prod_{i \in N} S_{i}}$ shows a strategy set of all players. A strategy profile $s\in S$ is a $n$-tuple vector $L=\{l^{tx}_{1},l^{tx}_{2},...,l^{tx}_{n}\}$ that shows numbers of TSCH Tx cells for all nodes in the network. $s$ can be presented by $(l^{tx}_{i},l^{tx}_{-i})_{i\in N}$ where $l^{tx}_{-i}=\{l^{tx}_{j}\in s| j\in N, j \neq i\}$. In our designed non-cooperative game, each IoT node $i \in N$ compete with other nodes for increasing the number of TSCH Tx timeslots to maximize its payoff. To inform children nodes of the number of TSCH Rx timeslots of the parent ($l^{rx}_{p_{i}}$), in GT-TSCH, we add one option filed to the structure of the DODAG Information Object (DIO) messages. Hence, when node $i$ broadcast a DIO message, it informs its children of $l^{rx}_{i}$.
\subsection{The Utility Function}
When a player chooses a strategy, the gained profit is determined by the utility function. IoT nodes compete with each other to increase their profit by allocating more TSCH Tx cells. As each TSCH Tx cell is used for forwarding only one data packet, the node that allocates a higher number of TSCH Tx cells, will have a higher bandwidth for the packet transmission. Different types of utility function (such as linear, exponential, and logarithmic) are commonly used in mathematical modeling of wireless networks. In GT-TSCH, we use a logarithmic utility function as it has the strict concavity property. The utility function of player $i$ is defined by 
\begin{equation}\label{eq:4.1}
u_{i}(l^{tx}_{i},l^{tx}_{-i})= (\overline{Rank_{i}})log(l^{tx}_{i}+1), \vspace{-1 mm}
\end{equation}\vspace{-1 mm}
where \vspace{-1 mm}
\begin{equation}\label{eq:4.1.1}
\overline{Rank_{i}}=\dfrac{MinStepofRank}{Rank_{i}-Rank_{min}}.
\end{equation}
$Rank_{i}$ is the Rank of node $i$ in the DAG topology, $Rank_{min}$ is the Rank of the root node, and $MinStepofRank$ is the minimum increase of the node's Rank at each hop. $u_{i}$ is designed in a way such that a player with shorter logical distance to the root node gains more profit. In other words, nodes with less Ranks have higher priorities in the cell allocation process. This strategy achieves load balancing in the DAG topology and reduces congestion by allocating more TSCH Tx cells to a parent node, compared to its children. 
\subsection{The Link Quality Cost Function}
Packet transmission is the main cause of energy consumption in low-power IoT networks. The quality of the wireless link that connects two adjacent neighbors has a significant impact on the number of retransmissions required for the successful delivery of a packet. When the quality of the wireless link is poor, allocating more TSCH Tx cells causes energy wastage due to redundant transmissions. This results in reducing node's lifetime that may lead to the disconnected network topology for nodes which are placed close to the border router. To cope with this problem, we consider the quality of the wireless link that connects a child node to its parent in a cost function by using the ETX metric. $ETX_{i,p_{i}}$ shows the quality of the wireless link which connects node $i$ to its parent $p_{i}$; it is estimated based on the Packet Reception Ratio (PRR) by \vspace{-3 mm}
\begin{equation}\label{eq:4.2}
ETX_{i,p_{i}}=\dfrac{1}{PRR_{i,p_{i}}} \ge 1.\vspace{-1 mm}
\end{equation}
When the level of noise or collisions on the wireless link is increased, GT-TSCH increases the cost of adding new TSCH Tx cells for saving energy resources consumed for retransmissions of collided packets. In our game model, the wireless link quality cost function of node $i$ is defined as\vspace{-1 mm}
\begin{equation}\label{eq:4.3}
d_{i}(l^{tx}_{i},l^{tx}_{-i})= l_{i}^{tx}(ETX_{i,p_{i}}-1).\vspace{-1 mm}
\end{equation}
By using this function in calculating the node's payoff, we reduce the node's incentive for receiving more TSCH Tx cells when the quality of the wireless link is degrading. This strategy avoids increasing the data packet generation rate for nodes placed in areas with high level of noise. \vspace{-1 mm}
\subsection{The Queue Cost Function}
Under heavy traffic loads, queue loss is the main cause of degrading PRR in low-power IoT networks \cite{que-rpl}. To overcome this problem, we consider having a high queue length as an indication of congestion in the design of our proposed TSCH scheduler. To avoid congestion, GT-TSCH monitors the node's queue and allocates more TSCH Tx cells for nodes with high queue length. To define a smooth queue metric which is resilient against the sudden changes, we use EWMA (Exponential Weighted Moving Average) to estimate the queue length periodically at the end of a time frame $t$ by\vspace{-1 mm}
\begin{equation}\label{eq:4.4}
Q_{i}(t)= \zeta(Q_{i}(t-1))+(1-\zeta)q_{i}(t),\vspace{-1 mm}
\end{equation}
where $\zeta$ is the smooth factor, $Q_{i}(t-1)$ is the weighted average queue metric of node $i$ at time frame $(t-1)$, and $q_{i}(t)$ is the number of queued packets at time frame $t$. To consider more priority for nodes with heavy load in the cell allocation process, we use this metric to decrease the cost of receiving more TSCH Tx cells for nodes with high queue lengths. By using the weighted average queue metric, we define the queue cost function of node $i$ as \vspace{-2 mm}
\begin{equation}\label{eq:4.5}
z_{i}(l^{tx}_{i},l^{tx}_{-i})= l_{i}^{tx}(1 - \dfrac{Q_{i}}{Q_{Max}}).\vspace{-1 mm}
\end{equation} 
In this function, by decreasing the node's load, we increase the queue cost to prioritize children nodes with high traffic loads in the cell allocation process. This strategy avoids congestion effectively by decreasing the queue loss and balancing the traffic load in the DAG topology.
\subsection{The Payoff Function}
We define the payoff function based on 1) the number of receiving TSCH Tx timeslots, 2) the place of a node in the DAG topology, 3) the estimated average queue length, and 4) the quality of the wireless link. The payoff function of node $i$ is defined as\vspace{-1 mm}
\begin{equation}\label{eq:4.6}
v_{i}(l^{tx}_{i},l^{tx}_{-i}) = \alpha u_{i}(l^{tx}_{i},l^{tx}_{-i}) - \beta d_{i}(l^{tx}_{i},l^{tx}_{-i}) - \gamma z_{i}(l^{tx}_{i},l^{tx}_{-i}),\vspace{-1 mm}
\end{equation} 
where $\alpha$, $\beta$, and $\gamma$ are user preference parameters for functions $u_{i}(l^{tx}_{i},l^{tx}_{-i})$, $d_{i}(l^{tx}_{i},l^{tx}_{-i})$, and $z_{i}(l^{tx}_{i},l^{tx}_{-i})$, respectively. The values of these parameters are set by considering the network topology and application features. An an example, for networks with high quality links under heavy traffic load, queue cost should have a higher priority in the payoff function compared to the link quality cost ($\gamma$ should be greater than $\beta$). Before finding the optimal value of $l^{tx}_{i}(\forall i \in N)$ for maximizing $v_{i}(l^{tx}_{i},l^{tx}_{-i})$, we prove the existence and uniqueness of the Nash equilibrium in the next subsection.\vspace{-1 mm}
\subsection{Proofs for the Existence and Uniqueness of Nash Equilibrium}\vspace{-1 mm}
In our defined game model, by finding the Nash equilibrium, each node can maximize its payoff by setting the number of TSCH Tx timeslots in the 6P \textit{ADD-REQUEST} message equal with the corresponding element in the Nash point. A strategy profile $s^{*}=[s_{1}^{*},s_{2}^{*},...,s_{n}^{*}] \in S$ is a Nash equilibrium when \vspace{-1 mm}
\begin{equation}\label{eq:4.7}
v_{i}(s^{*}_{i},s^{*}_{-i}) \geq v_{i}(s_{i},s^{*}_{-i}) \,\,\,\, \forall i \in N,\,\,\, s_{i},s^{*}_{i} \in S_{i}, \,\,\, s^{*}_{i} \neq s_{i}.
\end{equation}

\textbf{\it Theorem 1:} The existence of the Nash equilibrium is proved in our game model based on the Debreu's theorem \cite{debreu} as $\forall i \in N$, 1) $S_{i}$ is compact and convex, 2) $v_{i}(s_{i},s_{-i})$ is quasi-concave in $s_{i}$, 3) $v_{i}(s_{i},s_{-i})$ is continuous in $s_{-i}$ and $s_{i}$. $S_{i}$ is compact since it is defined in the range of $[l^{tx-min}_{i},l_{p_{i}}^{rx}]$ and $\forall x \in S_{i}, l^{tx-min}_{i} \le x \le l_{p_{i}}^{rx}$. $S_{i}$ is a convex set as for any $0 \leq \lambda \leq 1$ and $a,b \in S_{i}$, $\lambda a+(1-\lambda )b \in S_{i}$. The concavity of $v_{i}(s_{i},s_{-i})$  over $s_{i}$ is proved as its second partial derivative is negative, i.e.,\vspace{-1 mm}
\begin{equation}\label{eq:4.8}
\frac{\partial ^{2} v_{i}(s_{i},s_{-i})}{\partial s^{2}_{i}}=-\alpha\frac{\overline{Rank_{i}}}{(1+s_{i})^2} < 0.
\end{equation}
Finally, meeting the following conditions for $\forall s_{i} \in S_{i}$, $\forall s_{-i} \in \prod_{j\in N, j\ne i}S_{j}$ indicates that $v_{i}(s_{i},s_{-i})$ is continuous in both $s_{-i}$ and $s_{i}$: 1) $v_{i}(s_{i},s_{-i})$ is defined. 2) $\lim_{x \to s_{i}}v_{i}(x,s_{-i})$, $\lim_{y \to s_{-i}}v_{i}(s_{i},y)$ exist. 3) $\lim_{x \to s_{i}}v_{i}(x,s_{-i}) = \lim_{y \to s_{-i}}v_{i}(s_{i},y) = v_{i}(s_{i}, s_{-i})$.\vspace{2 mm}

\textbf{\it Theorem 2:} We prove the uniqueness of Nash equilibrium based on the Rosen's theorem for concave N-person games \cite{rosen} as 1) $\forall i \in N$, $S_{i}$ is closed, convex, and bounded (proved in theorem 1) and 2) $\forall s \in S$, the payoff functions $(v_{1},v_{2},...,v_{n})$ are diagonally strictly concave. The partial derivative of $v(s)$ respect to each variable $s_{i}(\forall i \in N)$ is defined as
\begin{equation}\label{eq:4.9}\vspace{-2 mm}
\nabla v(s)=[ \dfrac { \partial v_{1}(s_{1},s_{-1}) } { \partial s_{1} }, \dfrac { \partial v_{2}(s_{2},s_{-2}) } { \partial s_{2} }, ..., \dfrac { \partial v_{n}(s_{n},s_{-n}) } { \partial s_{n} } ]^{T}.
\end{equation}\vspace{1 mm}
The Jacobian matrix of $\nabla v(s)$ is defined by \vspace{-2 mm}
\begin{equation}\label{eq:4.10}
\begin{aligned}
J(\nabla v(s))= \hspace{150 mm}\\
\begin{bmatrix}
    \dfrac { \partial^{2} v_{1}(s_{1},s_{-1}) } { \partial s_{1}^2 }  & \dfrac { \partial^{2} v_{1}(s_{1},s_{-1}) } { \partial s_{1} \partial s_{2} } & \cdots & \dfrac { \partial^{2} v_{1}(s_{1},s_{-1}) } { \partial s_{1} \partial s_{n}} \\\ 
  \dfrac { \partial^{2} v_{2}(s_{2},s_{-2}) } { \partial s_{2} \partial s_{1} }  & \dfrac { \partial^{2} v_{2}(s_{2},s_{-2}) } { \partial s_{2}^2} & \cdots
   & \dfrac { \partial^{2} v_{2}(s_{2},s_{-2}) } { \partial s_{2} \partial s_{n}} \\\
 \vdots & \vdots & \ddots & \vdots \\\
  \dfrac { \partial^{2} v_{n}(s_{n},s_{-n}) } { \partial s_{n} \partial s_{1} }  & \dfrac { \partial^{2} v_{n}(s_{n},s_{-n}) } { \partial s_{n} \partial s_{2}} & \cdots
   & \dfrac { \partial^{2} v_{n}(s_{n},s_{-n}) } { \partial s_{n}^2}
  \end{bmatrix}. \hspace{80 mm}
\end{aligned}
\end{equation}
For any non-zero column vector $X \in \mathbb{R}^{n}$ and $\forall s \in S$, $X^T(J(\nabla v(s)) + J^{T}(\nabla v(s)))X<0$ that proves diagonal strict concavity of payoff functions.
\subsection{Game Solution}
We find optimal values of $l_{i}^{tx}(i\in N)$ to maximize the output of utility functions. The value of $l_{i}^{tx}$ must be set in the range of $[l^{tx-min}_{i},l_{p_{i}}^{rx}]$ to avoid congestion. Thus, finding the game solution can be modeled as a nonlinear programming \cite{convex} problem with two inequality constraints by 
\begin{equation}\label{eq:4.12}
\begin{array}{c@{\qquad}c@{\qquad}c}
\text{maximize} \;\; v_{i}(l^{tx}_{i},l^{tx}_{-i}) \\
\text{subject to:}\\
l^{tx-min}_{i}-l^{tx}_{i} \le 0, \; \; \;\;\;\; l^{tx}_{i} - l^{rx}_{p_{i}} \le 0.
\end{array} 
\end{equation}
We use the method of Lagrange multipliers \cite{lag1} to solve the optimization problem \eqref{eq:4.12}. In this method, a Lagrange function $\mathcal{L}_{i}(l^{tx}_{i},w_{1},w_{2})$ is defined by subtracting constraints as multiples of Lagrange multipliers; $w_{1}$ and $w_{2}$, from the objective function as \vspace{-1 mm}
\begin{equation}\label{eq:415}
\mathcal{L}_{i}(l^{tx}_{i},w_{1},w_{2})=v_{i}(l^{tx}_{i},l^{tx}_{-i})-w_{1}(l^{tx-min}_{i}-l^{tx}_{i})- w_{2}(l^{tx}_{i} - l^{rx}_{p_{i}}).
\end{equation} 
The solution of \eqref{eq:4.12} can be found by searching the value for $l_{i}^{tx}$ that satisfies all the following KKT conditions \cite{convex}:
\begin{enumerate}
\item{\it $l^{tx-min}_{i}-l^{tx}_{i} \le 0,\;\;\;\; l^{tx}_{i} - l^{rx}_{p_{i}} \le 0.$}\vspace{2 mm}
\item{\it $w_{1} \ge 0,\;\;\;\;  w_{2} \ge 0.$}\vspace{2 mm}
\item{$\frac {\partial v_{i}(l^{tx}_{i},l^{tx}_{-i})}{\partial l^{tx}_{i}}
-w_{1}\frac {\partial(l^{tx-min}_{i}-l^{tx}_{i})}{\partial l^{tx}_{i}}-w_{2}\frac {\partial(l^{tx}_{i} - l^{rx}_{p_{i}})}{\partial l^{tx}_{i}}=0.$}\vspace{2 mm}
\item{\it $w_{1}(l^{tx-min}_{i}-l^{tx}_{i})=0,\;\;\;\; w_{2}(l^{tx}_{i} - l^{rx}_{p_{i}})=0.$}\vspace{2 mm}
\end{enumerate}
Based on these conditions, the optimal solution for the number of TSCH Tx timeslots at node $i$ is found by
\begin{equation}\label{eq:416}
\begin{aligned}
l_{i}^{tx}=\hspace{80 mm} \\
\begin{cases}
l_{i}^{tx-min},&\hspace{-30 mm}\text{if}\;\;\;\;l_{i}^{tx-min} \hspace{-1 mm}\ge \hspace{-1 mm}(\frac{\alpha \overline{Rank_{i}}}{\gamma(1-\frac{Q_{i}}{Q_{max}}) + \beta(ETX_{i,p_{i}}-1)})-1,\\\\
l^{rx}_{p_{i}}, &\hspace{-30 mm}\text{if}\;\;\;\;l_{p_{i}}^{rx} \le (\frac{\alpha\overline{Rank_{i}}}{\gamma(1-\frac{Q_{i}}{Q_{max}}) + \beta(ETX_{i,p_{i}}-1)})-1,\\\\
(\frac{\alpha\overline{Rank_{i}}}{\gamma(1-\frac{Q_{i}}{Q_{max}}) + \beta(ETX_{i,p_{i}}-1)})-1, &\;\;\;\;\;\;\text{otherwise.}\\
\end{cases}
\end{aligned}
\end{equation}
Algorithm \ref{alg:2} shows the process of finding the optimal value of $l^{tx}_{i}$.

\begin{algorithm}[b]
\SetAlgoLined
\DontPrintSemicolon
Set values of $\alpha , \beta , \gamma$ \newline
Compute $X \gets (\frac{\alpha\overline{Rank_{i}}}{\gamma(1-\frac{Q_{i}}{Q_{max}}) + \beta(ETX_{i,p_{i}}-1)})-1$\newline
 \uIf{$l_{i}^{tx-min} \ge X $}
  {\vspace{3 mm}
  	   		$l_{i}^{tx} \gets l_{i}^{tx-min}$ \;
  }
  \Else
  {
  		\uIf{$l_{p_{i}}^{rx} \le X $}
        {\vspace{3 mm}
  	   		$l_{i}^{tx} \gets l_{p_{i}}^{rx}$ \;
        }
        \Else
        {
        $l_{i}^{tx} \gets X$
        }
  }
\caption{Computing the optimal value of $l_{i}^{tx}$ in the slotframe update process of GT-TSCH.}\label{alg:2}
\end{algorithm} \setlength{\textfloatsep}{3 mm}
\begin{table}[hb]
\caption{Contiki-NG configuration}
\label{table:table2}
\centering
\begin{tabular}{|*{3}{p{38 mm}|p{34 mm}|}}
\hline
TSCH Scheduling &GT-TSCH, Orchestra \\ \hline
TSCH timeslot length &15 milliseconds \\ \hline
Frequency hopping sequence &17, 23, 15, 25, 19, 11, 13, 21 \\ \hline
EB period &2s \\ \hline
Minimum DIO interval &300s\\ \hline
TSCH slotframe length &32 \\ \hline
Network layer   &uIPv6 + RPL \\ \hline
Objective function   &MRHOF \\ \hline
MAC layer &IEEE 802.15.4 + CSMA \\ \hline
Number of retransmissions &4 times \\ \hline 
\end{tabular}
\end{table} 
\section{Evaluation Results}\label{evaluation}
To examine the performance of our proposed method, we implement GT-TSCH on the Contiki-NG OS by using C programming language. Contiki-NG is an open source operating system designed for low-power, memory-constrained IoT devices. To make sure our contribution can be run on a wide range of IoT devices, we generate the executable binary code for Zolertia Firefly motes (Fig. {\ref{fig:zol}}) that have severe computational resource limitation (32 KB of RAM). The Zolertia Firefly mote is equipped with ARM Cortex-M3 CPU with 512KB Flash. Zolertia motes have been widely used to test and develop with the Open Thread project \cite{thread} released by Google. Moreover, To precisely estimate performance metrics such as radio duty cycle and end-to-end delay, we use Cooja, the Contiki network emulator. The unique characteristics of Cooja allow us to use the same binary code that we generated for Zolertia Firefly motes without making any modification. 

Among several TSCH schedulers, we compare performance of GT-TSCH with Orchestra \cite{orchestra} since 1) we could find the open-source implementation for Contiki released by authors \cite{ng2} and 2) it is the most well-known TSCH scheduler designed for low-power static IoT networks. In most of the recent proposed TSCH schedulers (e.g. \cite{ost, alice,a3, rea}), Orchestra was considered as the baseline for the performance evaluation. As it is shown in \cite{orchestra}, when the traffic load is light (1 packet per minute (ppm)), Orchestra can achieve 99\% packet delivery ratio. However, under heavy traffic (at least 30 ppm), its performance degrades dramatically. Unlike conventional networks, 30 ppm is considered a very heavy traffic load in low-power IoT networks. When the duration of a timeslot is 15 milliseconds, theoretically, delivering more than 66 packets per second (from all nodes) to a root node (receiver) is not possible ([1000/15]=66). We evaluate the performance of GT-TSCH in three types of scenarios: 1) increasing the packet generation rate from 30 to 165 packet per minute (ppm) on each IoT node, 2) increasing the size of the DODAG topology from 6 to 9 nodes, and 3) increasing the unicast slotframe length from 8 to 20. Table \ref{table:table2} shows the configuration of Contiki-NG OS. 
\begin{figure}[t] 	
 \centering
 \includegraphics[width=60 mm, height= 40 mm]{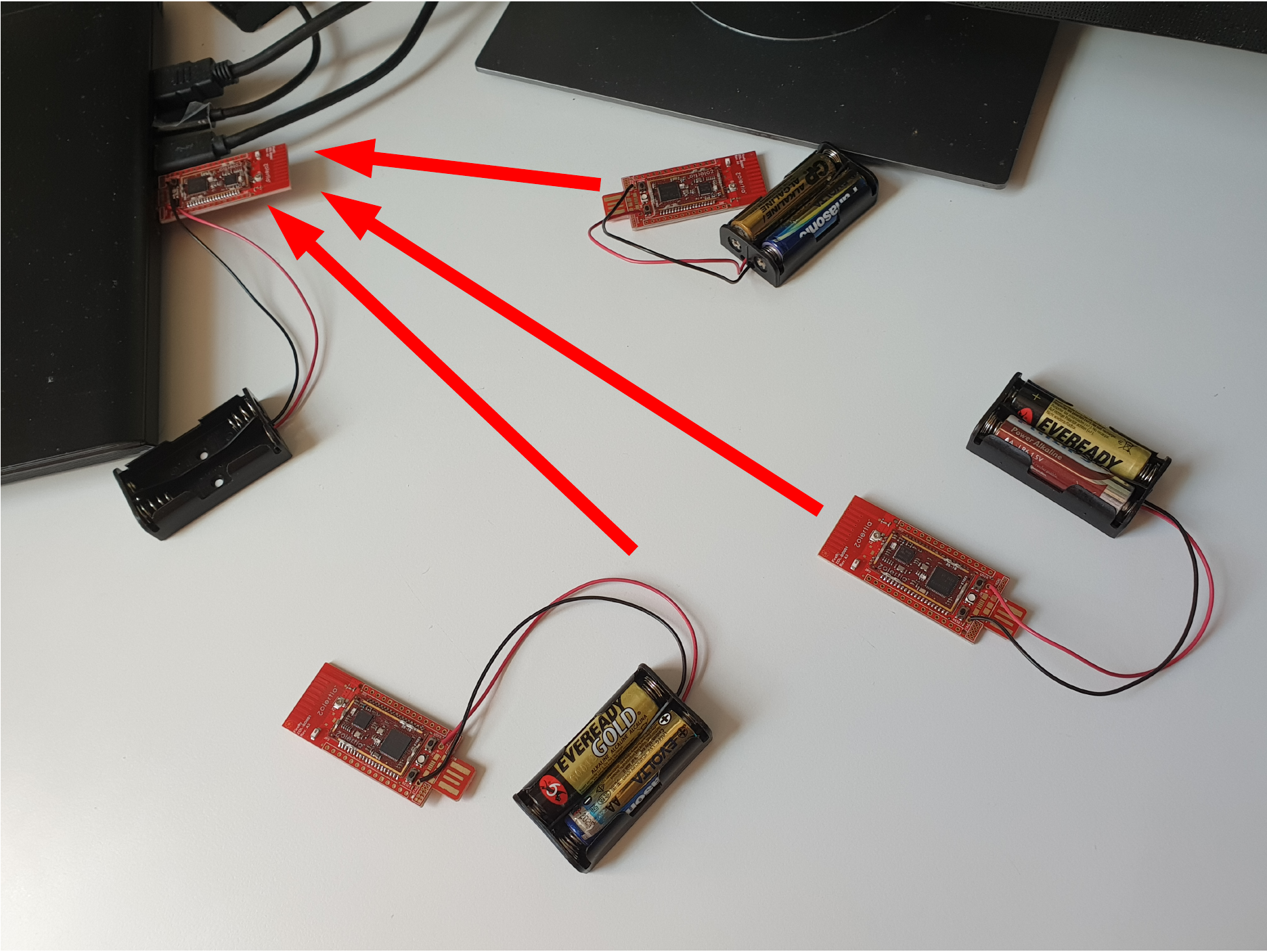}
 \caption{Zolertia Firefly IoT motes.}
 \label{fig:zol} 
\end{figure}

\begin{figure*}[ht]
  \centering
    \begin{subfigure}{0.32\textwidth}
 		\includegraphics[width=54 mm, height= 45 mm]{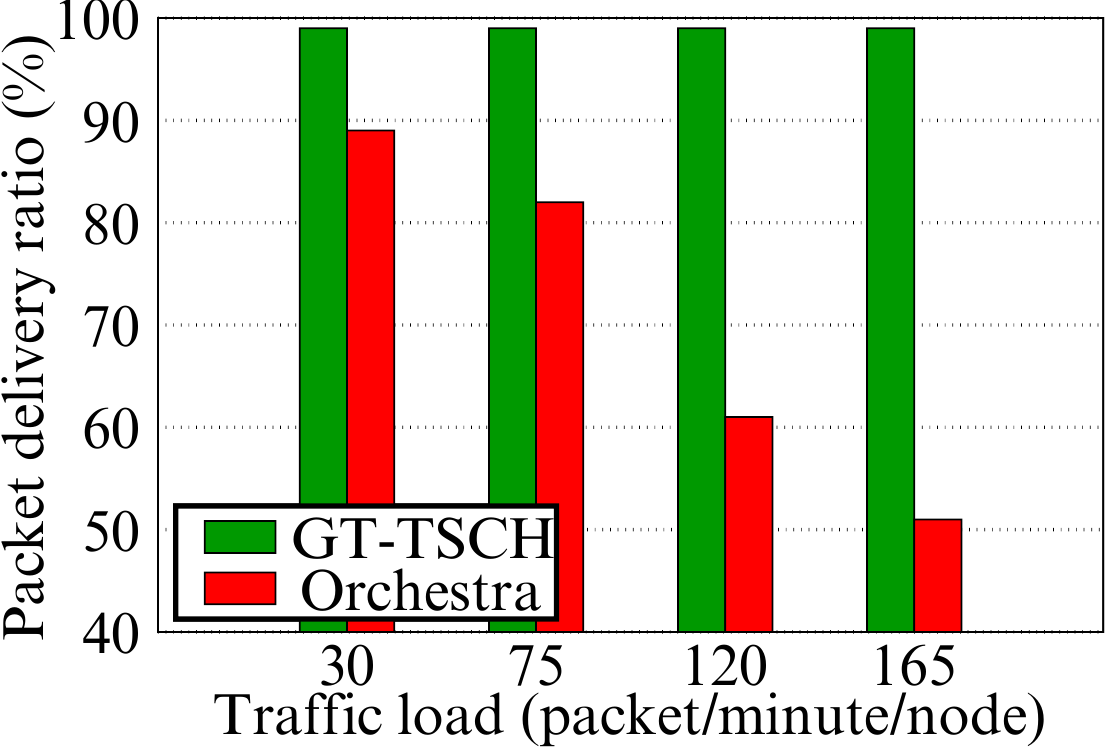} 
 		\caption{Packet delivery ratio}\vspace{5 mm}
 		\label{fig:e1a}
  	\end{subfigure}
  	\begin{subfigure}{0.32\textwidth}
    		\includegraphics[width=54 mm, height= 45 mm]{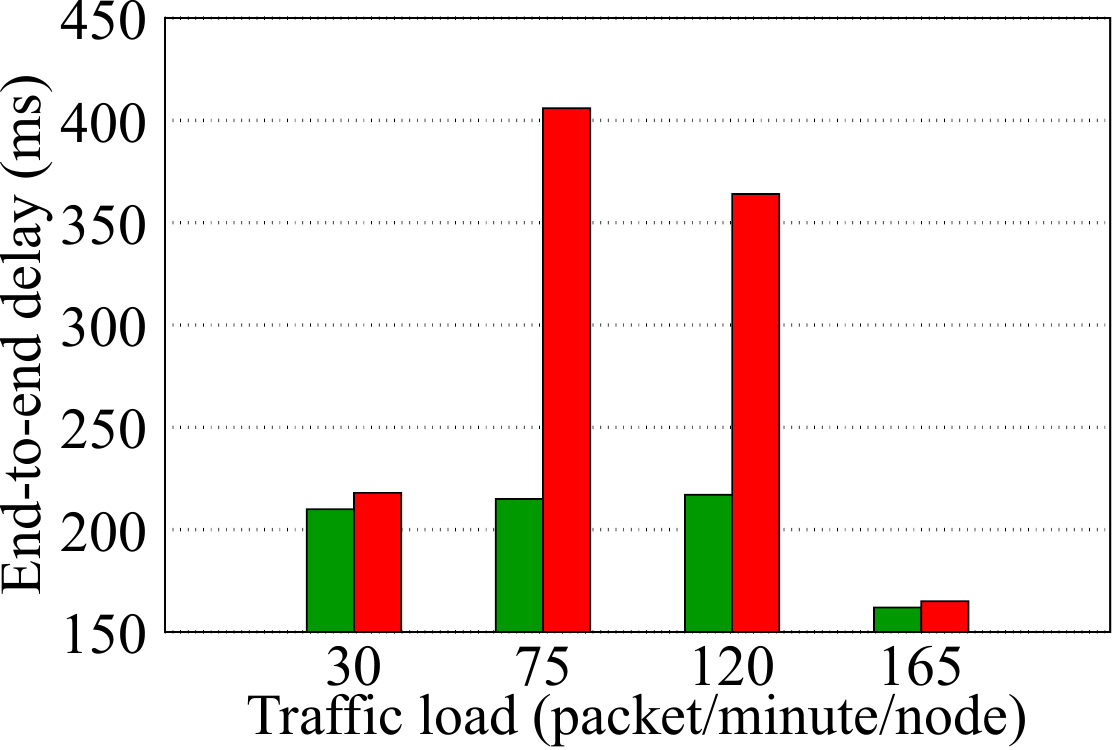}
    		\caption{Average end-to-end delay per packet}\vspace{5 mm}
    		\label{fig:e1b}
  	\end{subfigure}
  	\begin{subfigure}{0.32\textwidth}
    		\includegraphics[width=54 mm, height= 45 mm]{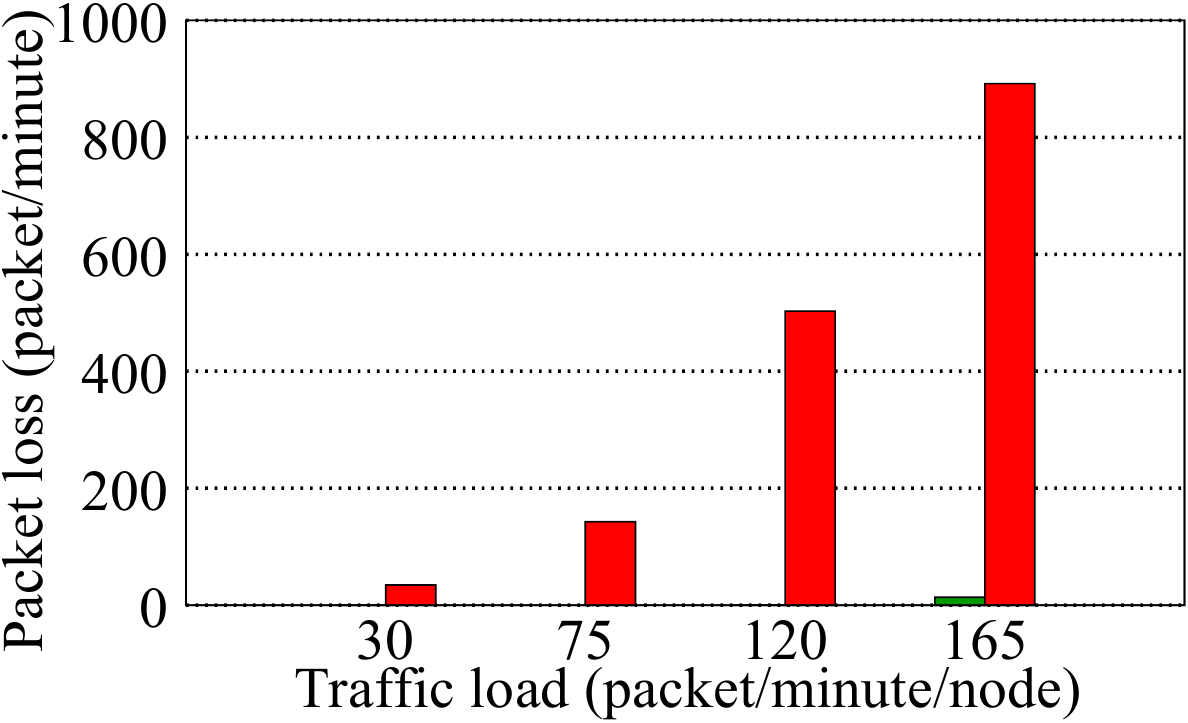}
    		\caption{Average number of lost packets}\vspace{5 mm}
    		\label{fig:e1c}
    \end{subfigure}
    \begin{subfigure}{0.32\textwidth}
    		\includegraphics[width=54 mm, height= 45 mm]{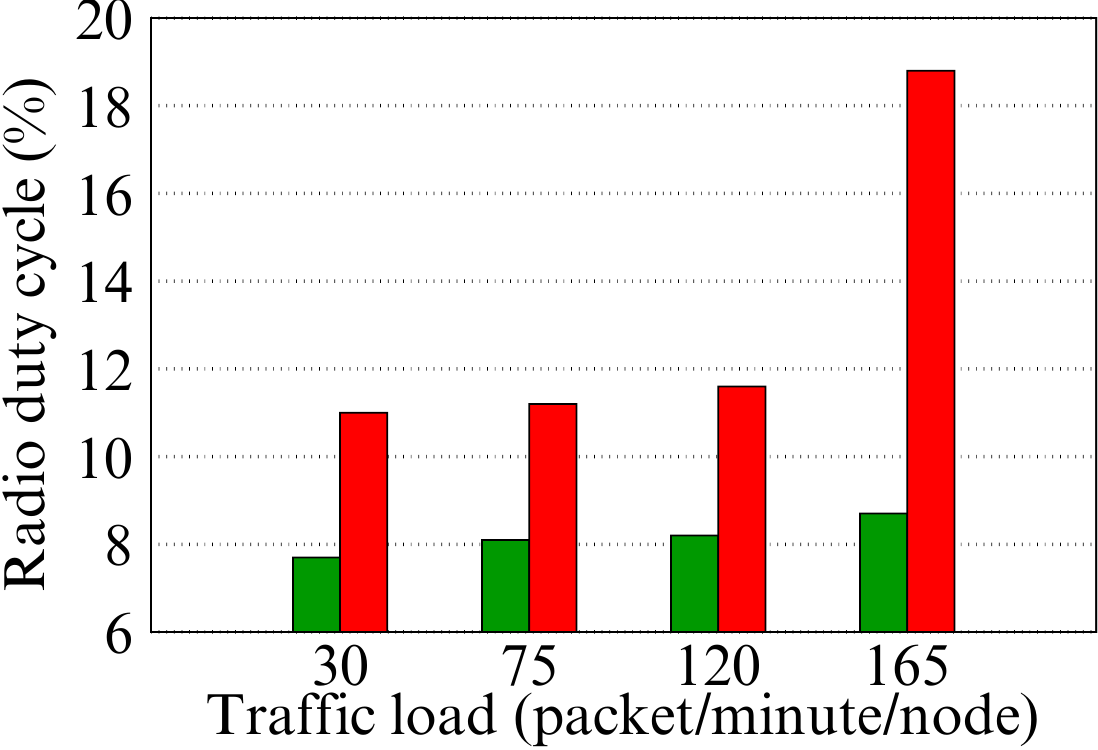} 
    		\caption{Average radio duty cycle per node}
    		\label{fig:e1d}
  	\end{subfigure}
  	\begin{subfigure}{0.32\textwidth}
 		\includegraphics[width=54 mm, height= 45 mm]{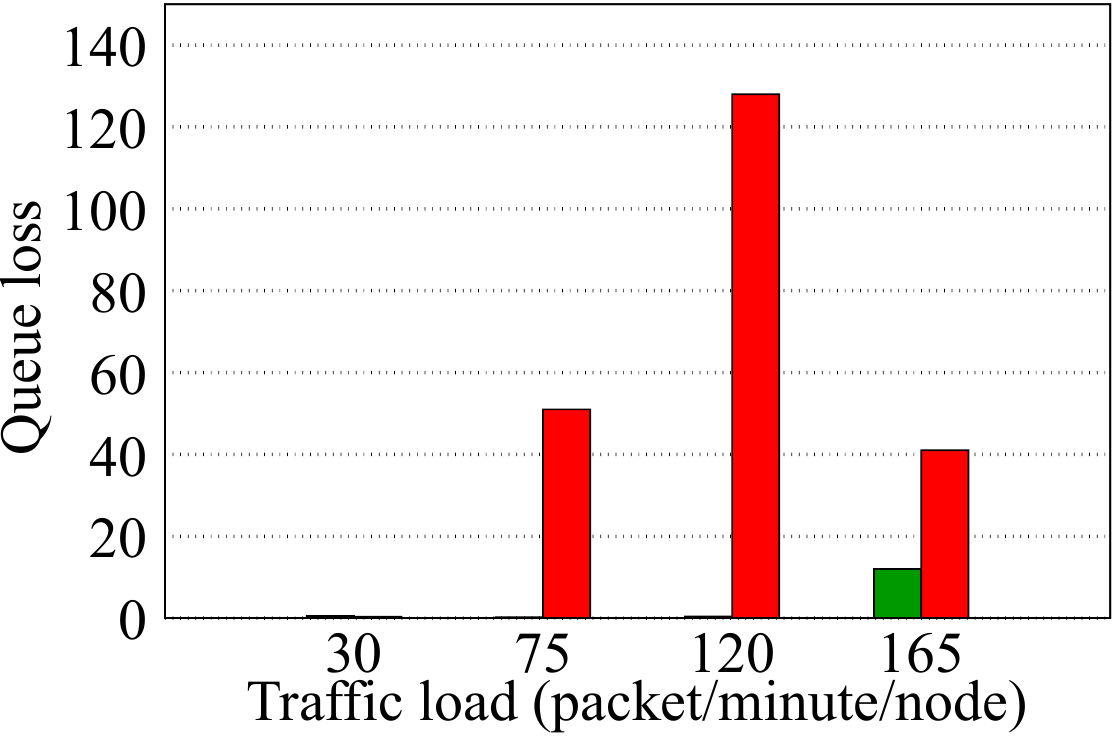}
 		\caption{Average queue loss per node}
 		\label{fig:e1e}
  	\end{subfigure}
    \begin{subfigure}{0.32\textwidth}
		\includegraphics[width=54 mm, height= 45 mm]{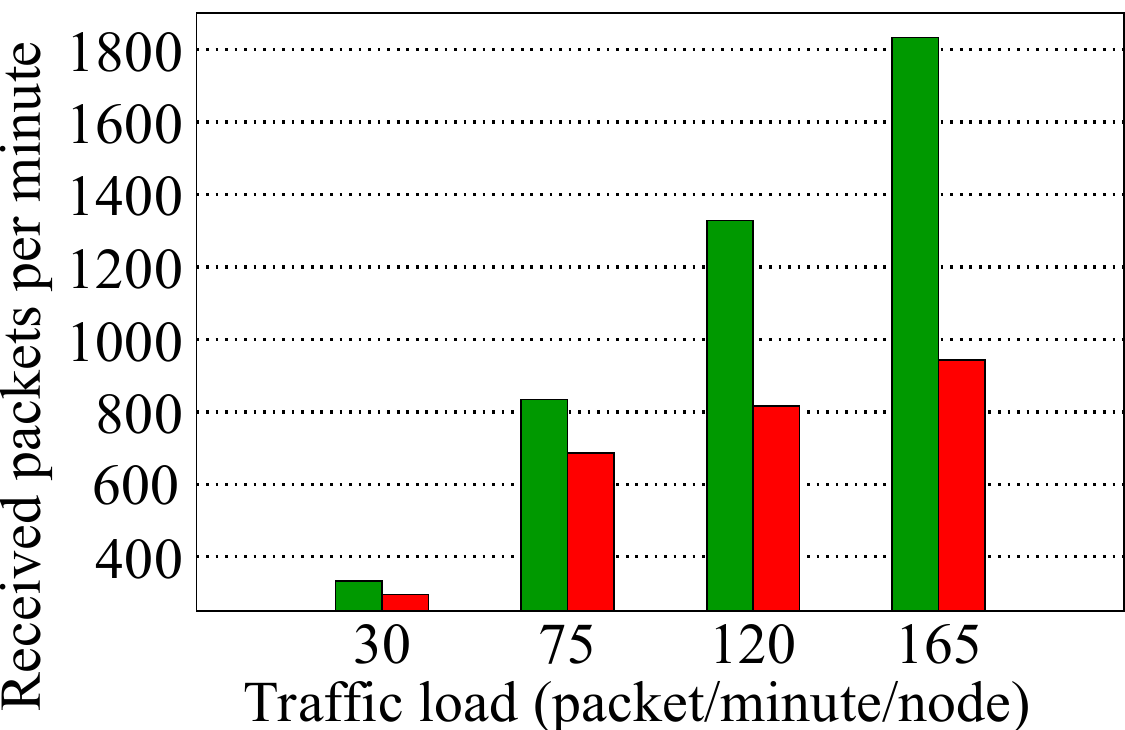}
		\caption{Average number of received packets}
		\label{fig:e1f}
    \end{subfigure}
 \caption{Performance of GT-TSCH and Orchestra according to traffic load.} 
\label{fig:12}
\end{figure*}

In the first set of our experiments, we evaluate the performance of GT-TSCH and Orchestra under varied traffic load in a network topology consists of two DODAGs with 14 nodes. Fig. {\ref{fig:e1a}} shows the ratio of packets delivered to root nodes. As shown in this figure, GT-TSCH keeps its PDR higher than 98\% by monitoring the queue length and updating the TSCH schedule for balancing the node's load. However, the performance of Orchestra dramatically decreased to around 50\% under high traffic load. For the traffic load with the rate of 165 ppm, the difference between performances of the two methods is considerably increased. GT-TSCH achieves 99\% PDR which is 45\% higher than that of Orchestra. Fig. {\ref{fig:e1b}} shows the performance results for the average end-to-end delay per packet. GT-TSCH has less end-to-end delay in all experiments as it reacts to the changes of data traffic quickly. By increasing the packet generation rate or the packet receiving rate, GT-TSCH adds more TSCH Tx timeslots into the slotframe. This strategy decreases the node's load and the average waiting time of packets in the queue. In addition, by monitoring the queue length, GT-TSCH assigns more TSCH Tx timeslots to nodes with high traffic loads. This strategy decreases the queue loss and accelerates the packet forwarding process. When the traffic load is 75 ppm, GT-TSCH has 215 milliseconds average end-to-end delay which is around 46\% less than that of Orchestra. For 165 ppm traffic load, the delay of both methods is reduced as they allocate many more TSCH timeslots for unicast transmission. 

Fig. \ref{fig:e1c}} shows the average number of lost packets per minute. As GT-TSCH delivers higher than 98\% of packets to root nodes, its average packet loss is less than one packet per minute for traffic loads with rates less than 120 ppm. Under high traffic load, at 165 ppm, GT-TSCH has the average of 13 lost ppm while the performance of Orchestra is degraded to 891 ppm as it does not take the node's load into account in the TSCH cell allocation process. To examine energy consumption of our proposed method, we estimate the radio duty cycle which is the percentage of time when the radio transmitter is on. The radio duty cycle is known as an effective criterion for monitoring energy resources in IoT networks. As shown in Fig. \ref{fig:e1d}}, Orchestra has a higher radio duty cycle compared to GT-TSCH as it cannot cope with wireless interference and collisions under high data traffic load. Without considering the node's load and the place of nodes in the DODAG topology for the frequency allocation process, Orchestra consumes more energy for retransmitting packets which are lost due to collisions. As a result, at 165 ppm, radio duty cycle of GT-TSCH is around 10\% less than that of Orchestra. Fig. {\ref{fig:e1e}} shows the impact GT-TSCH's load balancing mechanism on the average queue loss. GT-TSCH does not miss any packet when the data traffic rate is less than 165 ppm. Orchestra's queue lost is increased to around 130 packets at 120 ppm as it cannot allocate enough TSCH Tx timeslots for nodes which are placed close to root nodes in the DAG topology. The performance results for throughput are shown in Fig. {\ref{fig:e1f}}. By reducing the queue loss, and adapting the TSCH schedule dynamically based on the node's load, GT-TSCH can deliver more than 1800 packets to root nodes (at 165 ppm) which is around two times of Orchestra's throughput.
\begin{figure*}
  \centering
    \begin{subfigure}{0.32\textwidth}
 		\includegraphics[width=54 mm, height= 45 mm]{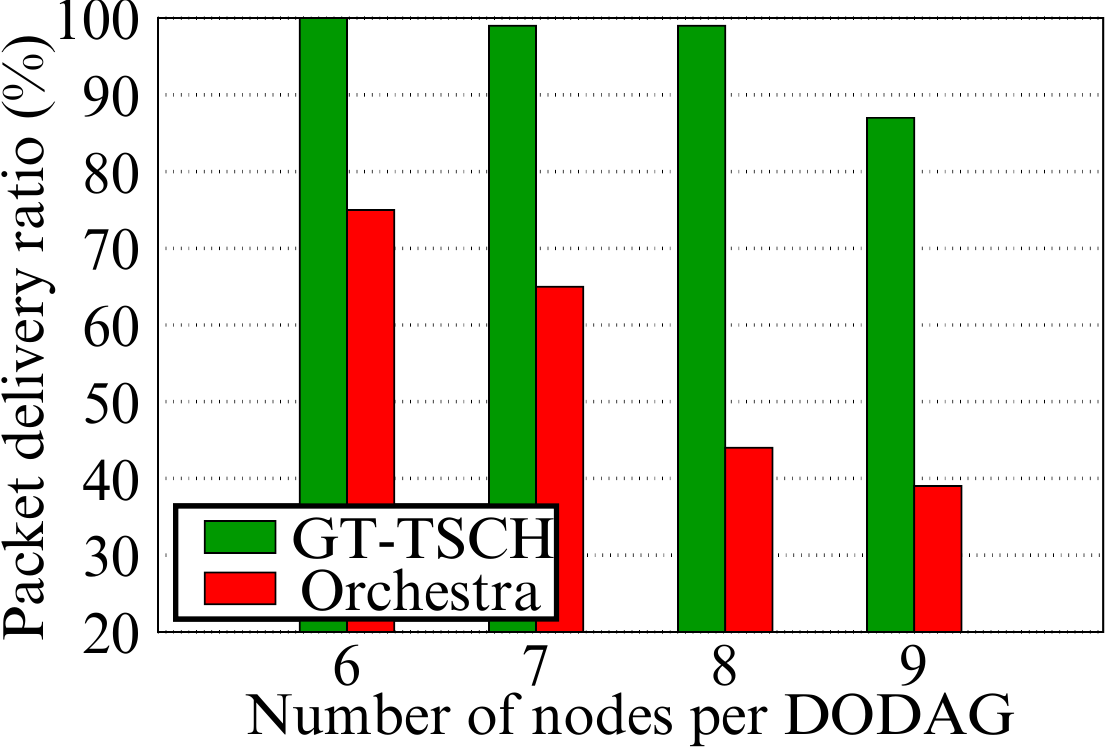} 
 		\caption{Packet delivery ratio}\vspace{5 mm}
 		\label{fig:e2a}
  	\end{subfigure}
  	\begin{subfigure}{0.32\textwidth}
    		\includegraphics[width=54 mm, height= 45 mm]{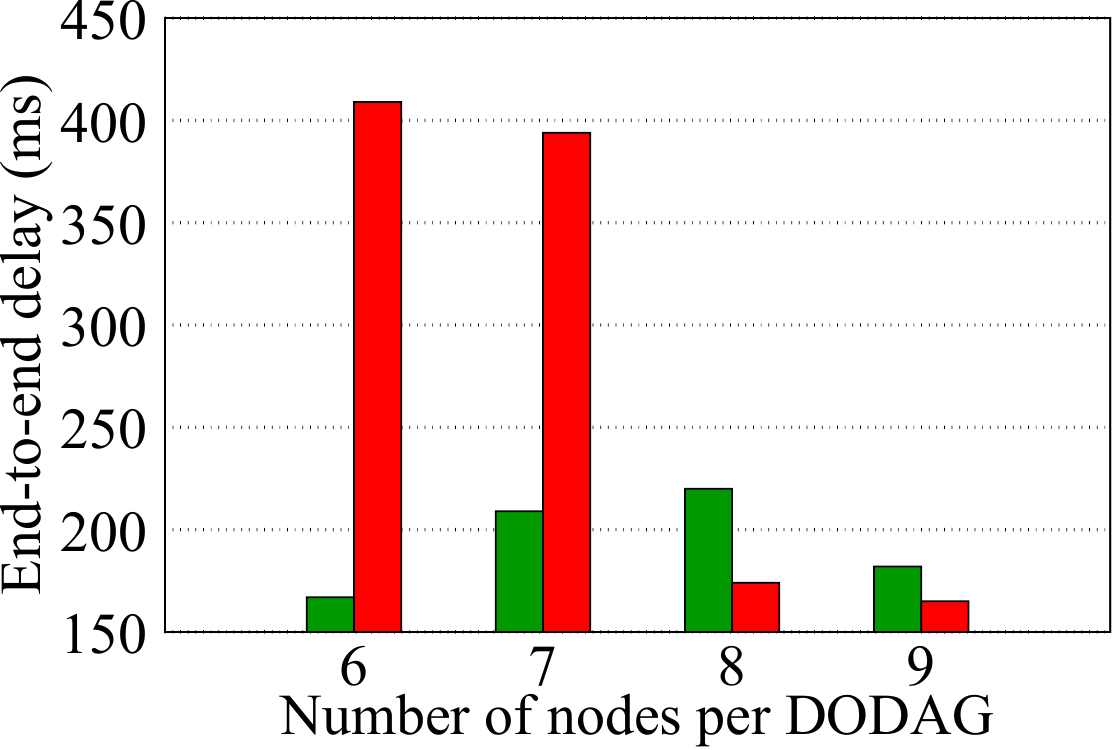}
    		\caption{Average end-to-end delay per packet}\vspace{5 mm}
    		\label{fig:e2b}
  	\end{subfigure}
  	\begin{subfigure}{0.32\textwidth}
    		\includegraphics[width=54 mm, height= 45 mm]{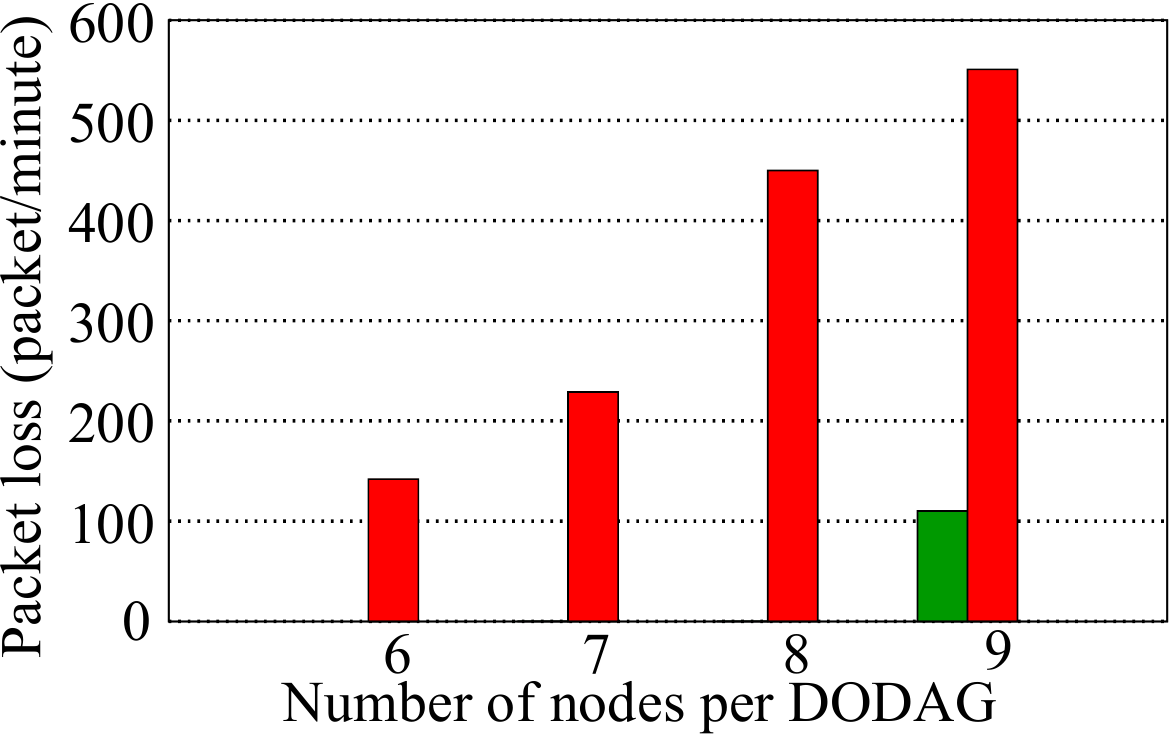}
    		\caption{Average number of lost packets}\vspace{5 mm}
    		\label{fig:e2c}
    \end{subfigure}
    \begin{subfigure}{0.32\textwidth}
    		\includegraphics[width=54 mm, height= 45 mm]{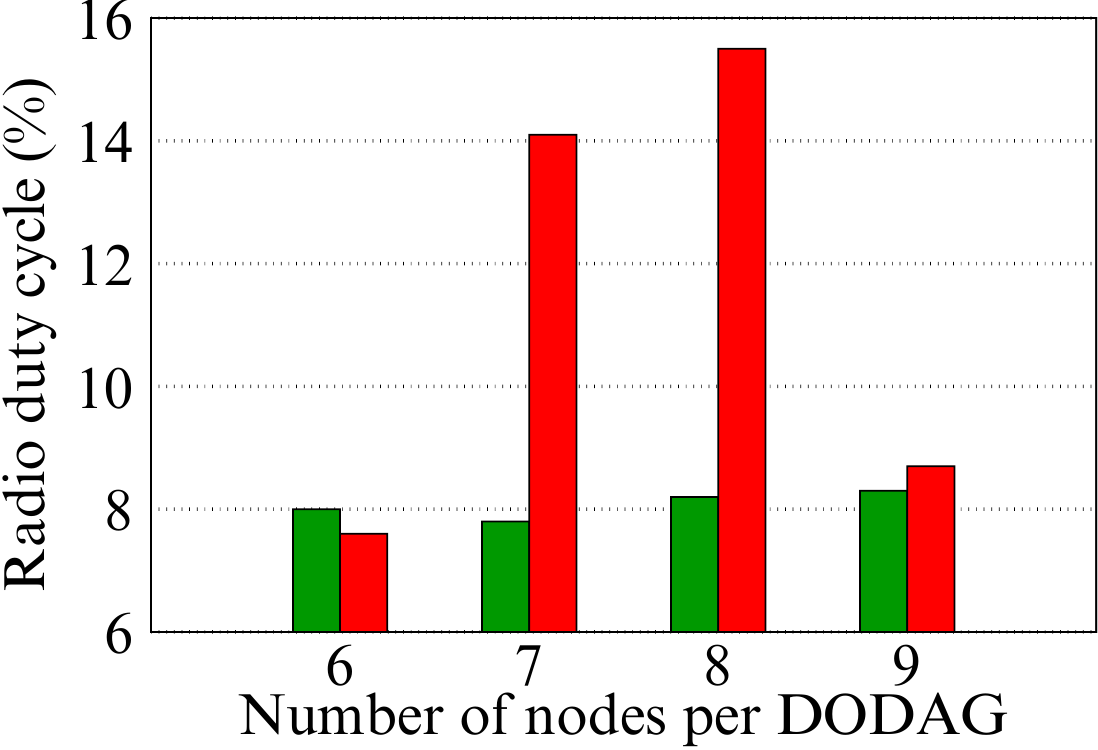} 
    		\caption{Average radio duty cycle per node}
    		\label{fig:e2d}
  	\end{subfigure}
  	\begin{subfigure}{0.32\textwidth}
 		\includegraphics[width=54 mm, height= 45 mm]{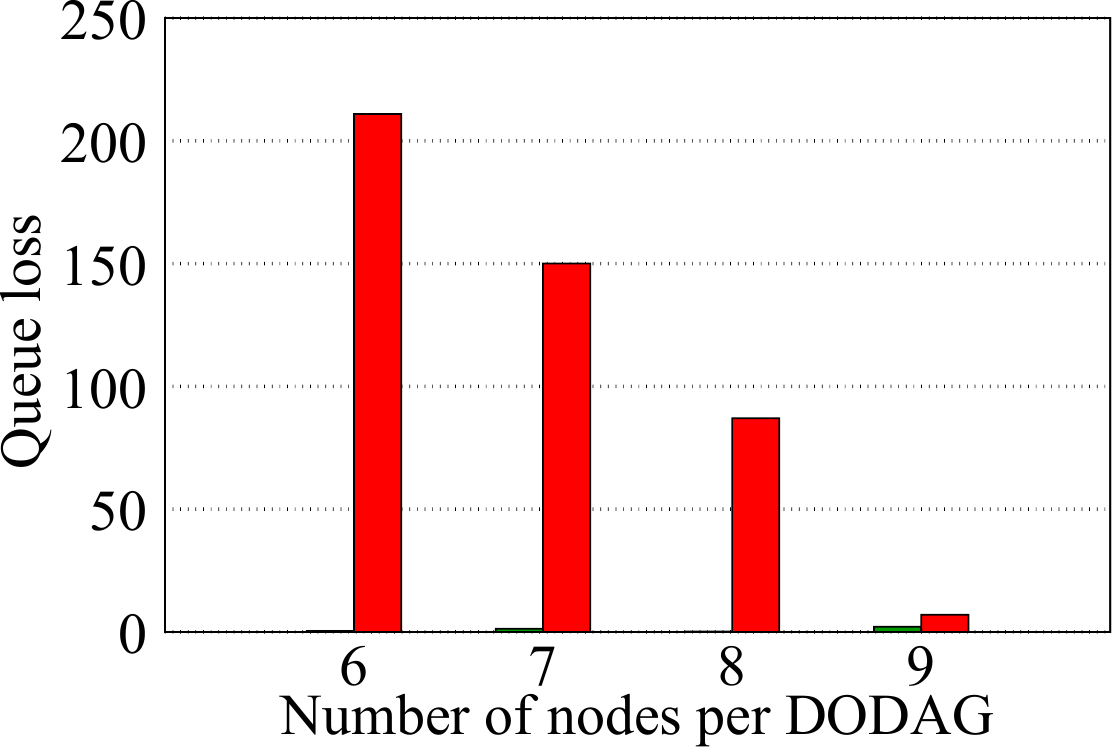}
 		\caption{Average queue loss per node}
 		\label{fig:e2e}
  	\end{subfigure}
    \begin{subfigure}{0.32\textwidth}
		\includegraphics[width=54 mm, height= 45 mm]{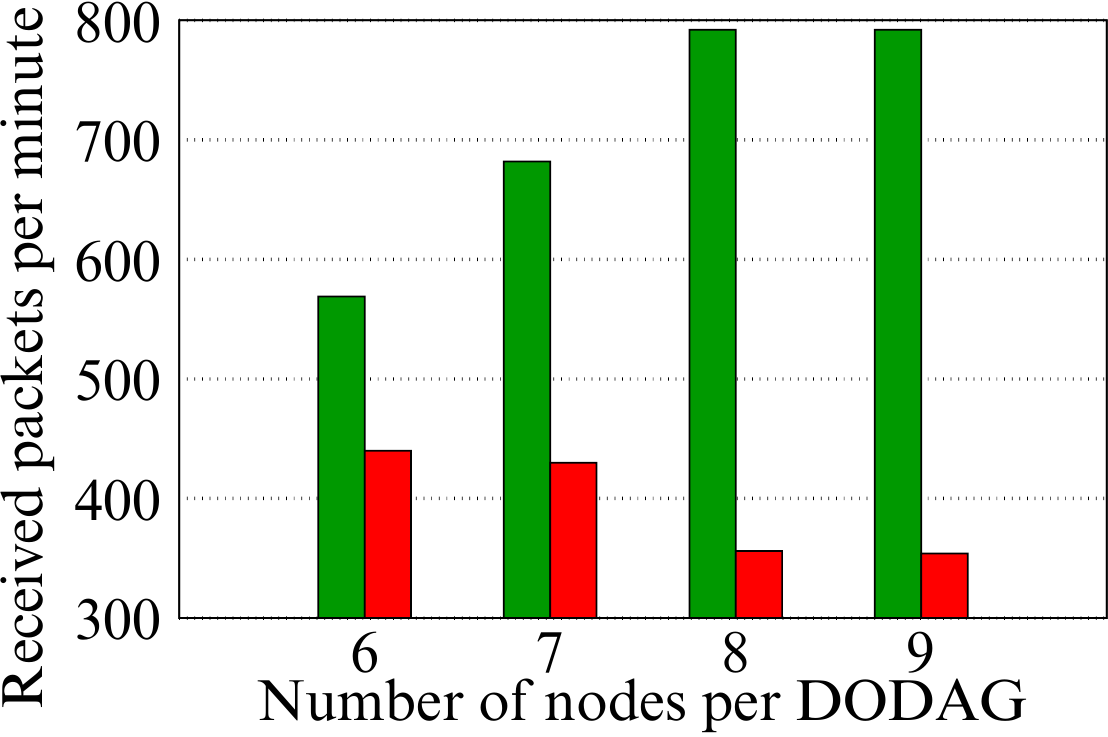}
		\caption{Average number of received packets}
		\label{fig:e2f}
    \end{subfigure}
 \caption{Performance of GT-TSCH and Orchestra according to the size of the DODAG topology.} 
\label{fig:e2}
\end{figure*}

In the second set of our experiments, we increase the number of nodes in a DODAG topology to examine the scalability of our contribution. Each DODAG topology is constructed by using only one root node. By adding more IoT nodes to the DODAG topology, we can find the maximum number of IoT nodes that TSCH scheduling algorithms support for each root node. For TSCH scheduling, the number of nodes per DODAG is a better criterion compared to the size of the network to examine the scalability. In many applications of LLNs (e.g., building automation), there is no common area in wireless ranges of DODAGs. Hence, expanding the network topology by adding more DODAGs does not change the performance results considerably. As an example, in a smart building, for each level, we have a DODAG that cannot be seen by IoT nodes placed in other levels. In these experiments, we increase the number of nodes per DODAG from 6 to 9. Thus, the total size of the network is increased from 12 to 18 nodes (for two DODAGs). We set the rate of traffic load at 120 ppm. As shown in Figs. {\ref{fig:e2a}}, {\ref{fig:e2c}}, and {\ref{fig:e2f}}, by increasing the DODAG size to more than 6 nodes, Orchestra's PDR is reduced since it cannot allocate enough TSCH timeslots for nodes with high traffic load. This results in increasing the packet loss to around three times when the DODAG size is increased from 6 to 8 nodes. On the other hand, GT-TSCH keeps its PDR higher than 98\% for up to 8 nodes per DODAG. When the DODAG size is 9, GT-TSCH cannot find free timeslots for the allocation of TSCH cells. Hence, its throughput is the same as that of a DODAG with 8 nodes (Fig. {\ref{fig:e2f}}). This leads to an increase in the packet loss and radio duty cycle (as shown in Figs. {\ref{fig:e2c}} and {\ref{fig:e2d}}). By increasing the DODAG size, the queue loss of Orchestra is reduced since it lost many packets due to lack of finding free TSCH Tx cells (Fig. {\ref{fig:e2e}}). In GT-TSCH, increasing the number of nodes results in allocating more TSCH Tx cells for packet forwarding in IoT nodes that play the roles of routers. This leads to a decrease in the average end-to-end delay, as it is shown in Fig. {\ref{fig:e2b}}.
\raggedbottom

\begin{figure*}[ht]
  \centering
    \begin{subfigure}{0.31\textwidth}
 		\includegraphics[width=54 mm, height= 45 mm]{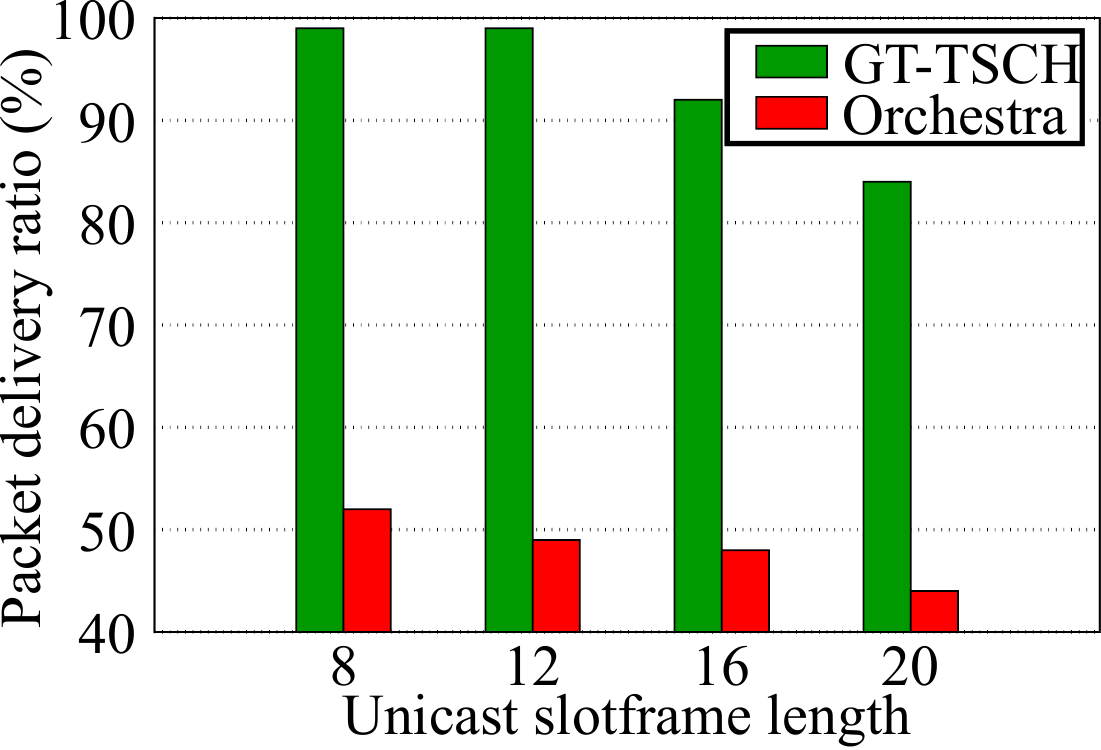} 
 		\caption{Packet delivery ratio}\vspace{5 mm}
 		\label{fig:e3a}
  	\end{subfigure}
  	\begin{subfigure}{0.32\textwidth}
    		\includegraphics[width=54 mm, height= 45 mm]{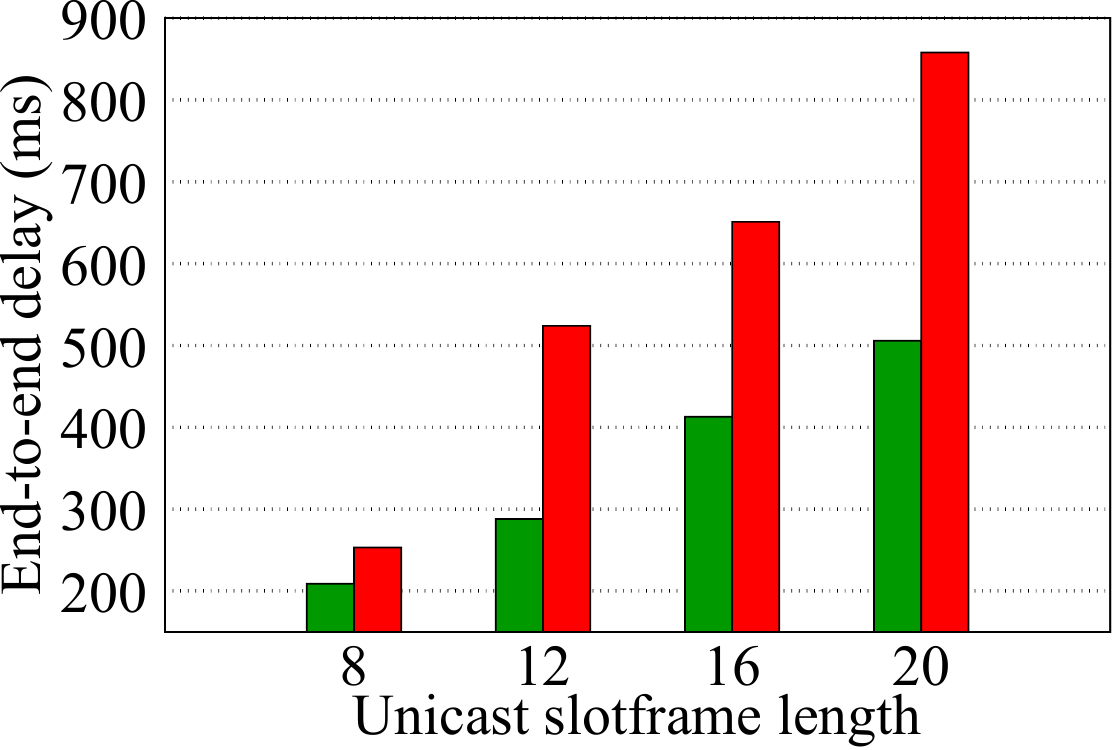}
    		\caption{Average end-to-end delay per packet}\vspace{5 mm}
    		\label{fig:e3b}
  	\end{subfigure}
  	\begin{subfigure}{0.32\textwidth}
    		\includegraphics[width=54 mm, height= 45 mm]{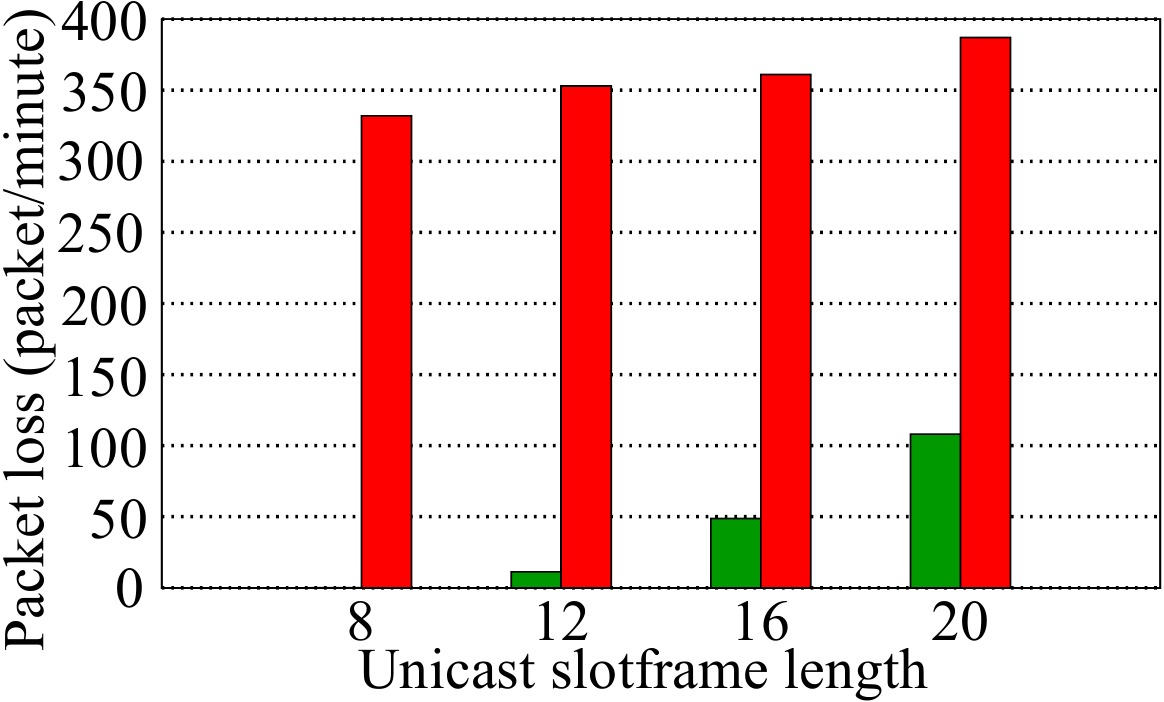}
    		\caption{Average number of lost packets}\vspace{5 mm}
    		\label{fig:e3c}
    \end{subfigure}
    \begin{subfigure}{0.32\textwidth}
    		\includegraphics[width=54 mm, height= 45 mm]{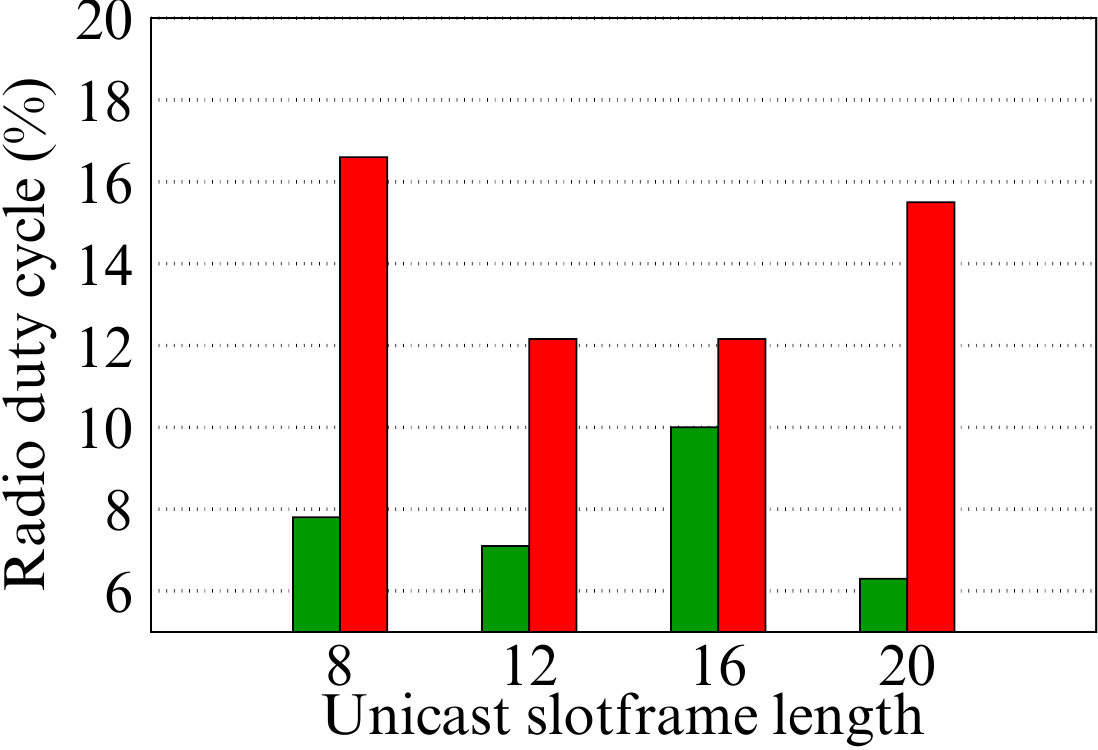} 
    		\caption{Average radio duty cycle per node}
    		\label{fig:e3d}
  	\end{subfigure}
  	\begin{subfigure}{0.32\textwidth}
 		\includegraphics[width=54 mm, height= 45 mm]{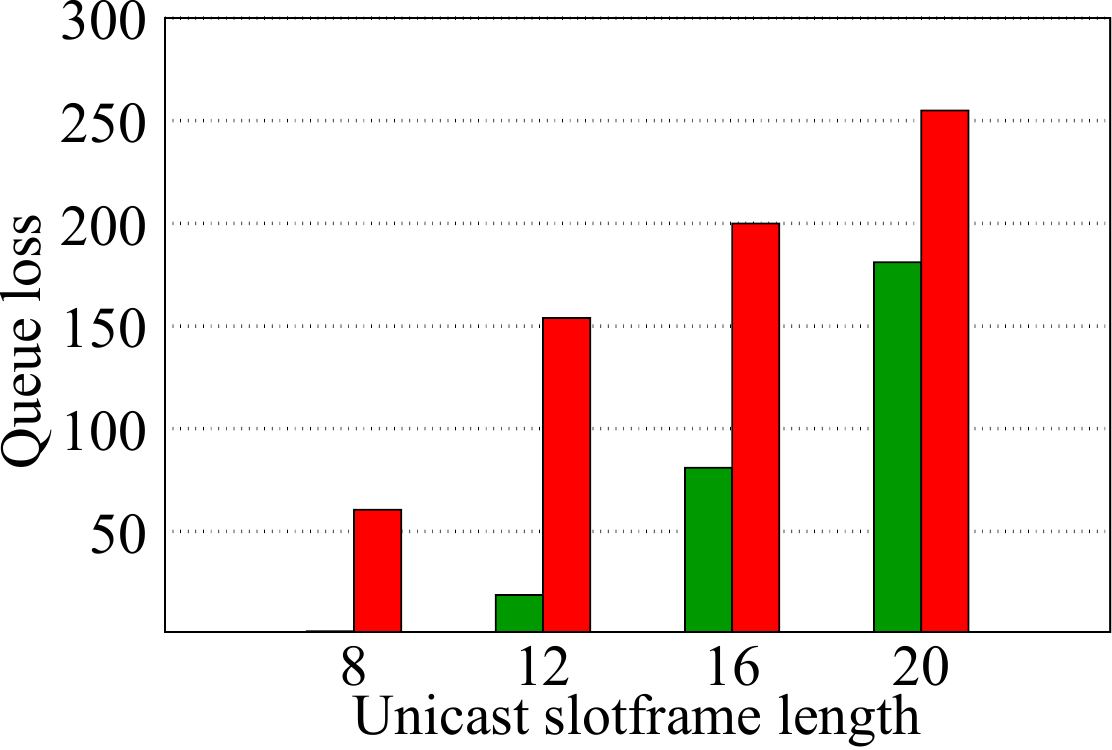}
 		\caption{Average queue loss per node}
 		\label{fig:e3e}
  	\end{subfigure}
    \begin{subfigure}{0.32\textwidth}
		\includegraphics[width=54 mm, height= 45 mm]{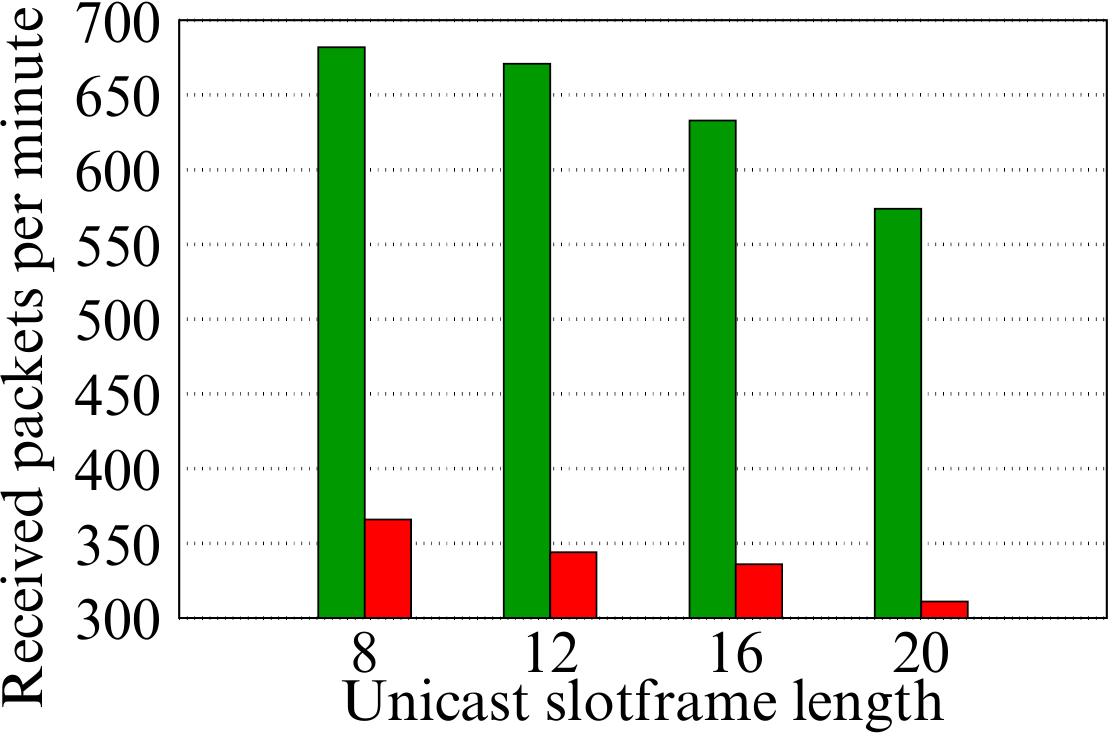}
		\caption{Average number of received packets}
		\label{fig:e3f}
    \end{subfigure}
 \caption{Performance of GT-TSCH and Orchestra for varied slotframe length.} 
\label{fig:e3}
\end{figure*}

The number of unicast timeslots in a slotframe has a significant impact on the performance of TSCH scheduling algorithms. In the third set of our experiments, we examine the performance of Orchestra and GT-TSCH for the slotframe with different sizes. GT-TSCH uses only one slotframe for all packet transmissions while Orchestra uses different slotframes for various types of traffic (e.g RPL, TSCH, and application data). To have a fair evaluation in these experiments, we set the size of the GT-TSCH's slotframe equal to four times of the unicast slotframe size of Orchestra. As Fig. {\ref{fig:e3a}} shows, GT-TSCH keeps its PDR higher than 80\% in all experiments while the PDR of Orchestra is reduced to less than 50\% when the slotframe size is higher than 8. This leads to an increase in the average delay by more than two times (Fig. {\ref{fig:e3b}}) and keeping the radio duty cycle higher than 12\% (Fig. {\ref{fig:e3d}}). By prioritizing children nodes with high traffic load in the timeslot allocation process, GT-TSCH keeps its throughput higher than 550 ppm in all experiments. By taking the queue length into account and avoiding collisions in assigning frequency channels to children nodes, GT-TSCH reduces the queue loss and packet loss to less than 200 packets and 110 ppm, respectively (Figs. {\ref{fig:e3e}} and {\ref{fig:e3c}}).
\section{Conclusion}\label{conclusion}
In this paper, we introduced GT-TSCH, a distributed dynamic TSCH scheduler designed based on the non-cooperative game theory for low-power IoT networks. GT-TSCH adapts the TSCH schedule by monitoring 1) the queue length, 2) the rates of packet generation and packet forwarding, 3) the quality of wireless links, and 4) the place of the node in the DAG topology. By using a lightweight method for frequency allocation, GT-TSCH avoids collisions and wireless interference. By considering selfish behavior of nodes in packet forwarding, GT-TSCH finds the optimal number of TSCH Tx cells for updating the TSCH schedule. To examine the performance of our proposed method, we implemented GT-TSCH on Zolertia Firefly IoT motes and the Contiki-NG OS. Evaluation results revealed that our contribution can improve PDR and reduce latency compared to the state-of-the-art method.
\bibliography{gt-tsch.bib}
\bibliographystyle{IEEEtran}

\end{document}